\documentclass{IEEEtran}
\usepackage{amsfonts}
\usepackage{mathrsfs}
\usepackage{amssymb,amsmath}
\usepackage[noblocks]{authblk}
\usepackage[compress]{cite}
\usepackage{bm}
\usepackage{algorithm}
\usepackage{algorithmic}
\usepackage{epsfig,graphics,subfigure}
\usepackage{theorem}
\usepackage{array,color}
\usepackage[compress]{cite}
\usepackage[OT1]{fontenc}

\usepackage[dvipsnames]{xcolor}

\allowdisplaybreaks
\theoremheaderfont{\normalfont\bfseries}

\newcommand{\T}{\mathrm{T}}
\newcommand{\vecm}[1]{\mathrm{vec}{#1}}

\begin{document}

\title{\LARGE A Framework on Hybrid MIMO Transceiver Design based on Matrix-Monotonic Optimization}

\author{
	\normalsize Chengwen Xing, \textit{Member, IEEE,} Xin Zhao, Wei Xu, \textit{Senior Member, IEEE,}\\ Xiaodai Dong, \textit{Senior Member, IEEE,} and Geoffrey Ye Li, \textit{Fellow, IEEE}
	
	\thanks{
		C. Xing and X. Zhao are with the School of Information and Electronics, Beijing Institute of Technologies, Beijing 100081, China. (e-mail: chengwenxing@ieee.org and xinzhao.eecs@gmail.com).

		W. Xu is with the National Mobile Communications Research Lab, Southeast University, Nanjing 210096, China. He is also with the Department of ECE, University of Victoria, Victoria BC, Canada. (email: wxu@seu.edu.cn).

		X. Dong is with the Department of Electrical and Computer Engineering, University of Victoria, Victoria, BC V8W 3P6, Canada. (e-mail: xdong@ece.uvic.ca).

		G. Y. Li is with the School of Electrical and Computer Engineering, Georgia Institute of Technology, Atlanta 30332-0250, GA USA (e-mail: liye@ece.gatech.edu).
	}
}

\maketitle

\begin{abstract}
	\normalsize
	Hybrid transceiver can strike a balance between complexity and performance of multiple-input multiple-output (MIMO) systems.
	In this paper, we develop a unified framework on hybrid MIMO transceiver design using matrix-monotonic optimization.
	The proposed framework addresses general hybrid transceiver design, rather than just limiting to certain high frequency bands, such as millimeter wave (mmWave) or terahertz bands or relying on the sparsity of some specific wireless channels.
	In the proposed framework, analog and digital parts of a transceiver, either linear or nonlinear, are jointly optimized.
	Based on matrix-monotonic optimization, we demonstrate that the combination of the optimal analog precoders and processors are equivalent to eigenchannel selection for various optimal hybrid MIMO transceivers.
	From the optimal structure, several effective algorithms are derived to compute the analog transceivers under unit modulus constraints.
	Furthermore, in order to reduce computation complexity, a simple random algorithm is introduced for analog transceiver optimization.
	Once the analog part of a transceiver is determined, the closed-form digital part can be obtained.
	Numerical results verify the advantages of the proposed design.
\end{abstract}


\section{Introductions}
\label{sect:intro}

The great success of multiple-input multiple-output (MIMO) technology makes it widely accepted for current and future high data-rate communication systems\cite{PimmWaveIntroMCOMM2011}. Acting as a pillar to satisfy data hungry applications, a natural question is how to reduce the cost of MIMO technology, especially that of large scale antenna arrays. The traditional setting of one radio-frequency (RF) chain per antenna element is too expensive for large-scale MIMO systems, especially at high frequencies, such as millimeter wave bands or Terahertz bands. Hybrid analog-digital architecture is promising to alleviate the straits and strike a balance between the cost and the performance of practical MIMO systems.

A typical hybrid analog-digital MIMO transceiver consists of four components, i.e., digital precoder, analog precoder, analog processor, and digital processor \cite{HybSurvey2017}. In the early transceiver design, hybrid MIMO technology is often referred to antenna selection \cite{SwitchTSP2003, SwitchTWC2005} to reap spatial diversity. In these works, analog switches are used in the radio-frequency domain. Phase-shifter based soft antenna selection \cite{VariablePhaseTSP2005,AntennaSeLTWC2006,SwitchAnalyzeTVT2009} has been proposed to improve performance for correlated MIMO channels. Nowadays, the phase-shifter based hybrid structure has been widely used.

For a phase shifter, only signal phase, instead of both magnitude and phase, can be adjusted. Thus, the optimization of a MIMO transceiver with phase shifters becomes complicated due to the constant-modulus constraints on analog precoder and analog processor.
It has been shown in \cite{YPLinTSP2017} that the performance of a full-digital system can be achieved when the number of shifters is doubled in a phase-shifter based hybrid structure. However, this can hardly be practical due to the requirement on a large number of phase shifters, especially in large-scale MIMO systems. As a matter of fact, the phase shifters in large-scale MIMO systems have been considered to be a burden sometimes. Thus, sub-connected hybrid structure has emerged as an alternative option \cite{CLinJSAC2017,dynamicSubTWC2017} and it has received much attention recently \cite{CLinJSAC2017, CLinWCNC2017,OverlapLSP2017,dynamicSubTWC2017,HybridSubarrayTCOMM2018,ReconfigureTCOMM2018}.

Unit modulus and discrete phase make the optimization of analog transceivers nonconvex and thus difficult to address \cite{SwitchTSP2003,SwitchTWC2005}. There have been some works on hybrid transceiver optimization considering different design limitations and requirements. Their motivation is to exploit the underlying structures of the hybrid transceiver to achieve high performance but with low complexity.

Early hybrid transceiver design is based on approximating digital transceivers in terms of the norm difference between all-digital design and the hybrid counterpart. For the millimeter wave (mmWave) band channels, which are usually with sparsity, an orthogonal matching pursuit (OMP) algorithm has been used in signal recovery for the hybrid transceiver \cite{HeathTWC2014}. In order to overcome the non-convexity in hybrid transceiver optimization, some distinct characteristics of mmWave channels must be exploited \cite{HeathTWC2014}. This methodology is a compromise on the constant-modulus constraint, which has been validated in different environments, including multiuser and relay scenarios \cite{hybridMUTVT2015,hybridRelay2014}. However, it has been found later on that the OMP algorithm cannot achieve the optimal solution sometimes. A singular-value-decomposition (SVD) based descent algorithm \cite{DongTSP2017} has been proposed, which is nearly optimal. An alternative fast constant-modulus algorithm \cite{fastAlgTSP2017} has also been developed to reduce the gap between the analog and digital precoders. The above methods are hard for complex scenarios due to high computation complexity \cite{HanzoHybridTCOMM2016,LLDaiJSAC2016,PhaseDirectTCOMM2017}. Therefore, based on the idea of unitary matrix rotation, several algorithms \cite{MaiRuikaiTWC,Y.C.Eldar2017} have been proposed to improve the approximation performance while maintaining a relative low complexity at the same time.

On the other hand, some works for hybrid precoding design are based on codebooks, which relax the problem into a convex optimization problem \cite{JHWangTSP2017}. However, the codebook-based algorithm suffers performance loss if channel state information (CSI) is inaccurate \cite{ImpactTSP2016}. In order to reduce the complexity of codebook design and the impact of partial CSI, special structures of massive MIMO channels \cite{WeiyuHybridJSTSP2016,WeiyuHybridOFDMJSAC2017}, can be exploited. Recently, an angle-domain based method has been proposed from the viewpoint of array signal processing \cite{LinHaiJSAC,FFGaoTWC2017angle}, which provides a useful insight on hybrid analog and digital signal processing. Based on the concept of the angle-domain design, some mathematical approaches, such as matrix decomposition algorithm, have been developed \cite{HybridKroneckerJSAC2017,LowComplexityTWC2018}. Energy efficient hybrid transceiver design for Rayleigh fading channels has been investigated in \cite{PayamiHybridTWC2016}. Hybrid transceiver optimization with partial CSI and  with discrete phases has been discussed in \cite{ANovelTCOMM2017} and \cite{DiscreteTVT2017}, respectively.

Hybrid MIMO transceivers are not only limited to mmWave frequency bands or terahertz frequency bands but also potentially work in other frequency bands. The transceiver itself could either be linear or nonlinear. Moreover, the performance metrics for MIMO transceiver could be different, including capacity, mean-squared error (MSE), bit-error rate (BER), etc. A unified framework on hybrid MIMO transceiver optimization will be of great interest. In this paper, we will develop a unified framework for hybrid linear and nonlinear MIMO transceiver optimization. Our main contributions are summarized as follows.
\begin{itemize}

\item  Both linear and nonlinear transceivers with Tomlinson-Harashima precoding (THP) or deci-sion-feedback detection (DFD) are taken into account in the proposed framework for hybrid MIMO transceiver optimization.

\item Different from the existing works in which a single performance metric is considered for hybrid MIMO transceiver designs, more general performance metrics are considered.

\item Based on matrix-monotonic optimization framework, the optimal structures of both digital and analog transceivers with respect to different performance metrics have been analytically derived. From the optimal structures, the optimal analog precoder and processor correspond to selecting eigenchannels, which facilitates the analog transceiver design. Furthermore, several effective analog design algorithms have been proposed.
\end{itemize}

The rest of this paper is organized as follows. In Section II, a general hybrid system model and the MSE matrices corresponding to different transceivers are introduced. In Section III, a unified hybrid transceiver is discussed in detail and the related transceiver optimization is present. In Section IV, the optimal structure of digital transceivers is derived based on matrix-monotonic optimization. In Section V, basic properties of the optimal analog precoder and processor are investigated, based on which effective algorithms to compute the analog transceiver are proposed. Next, in Section VI, simulation results are provided to demonstrate the performance advantages of the proposed algorithms. Finally,
\begin{figure*}[!ht]
	\centering
	\includegraphics[width = 1\textwidth]{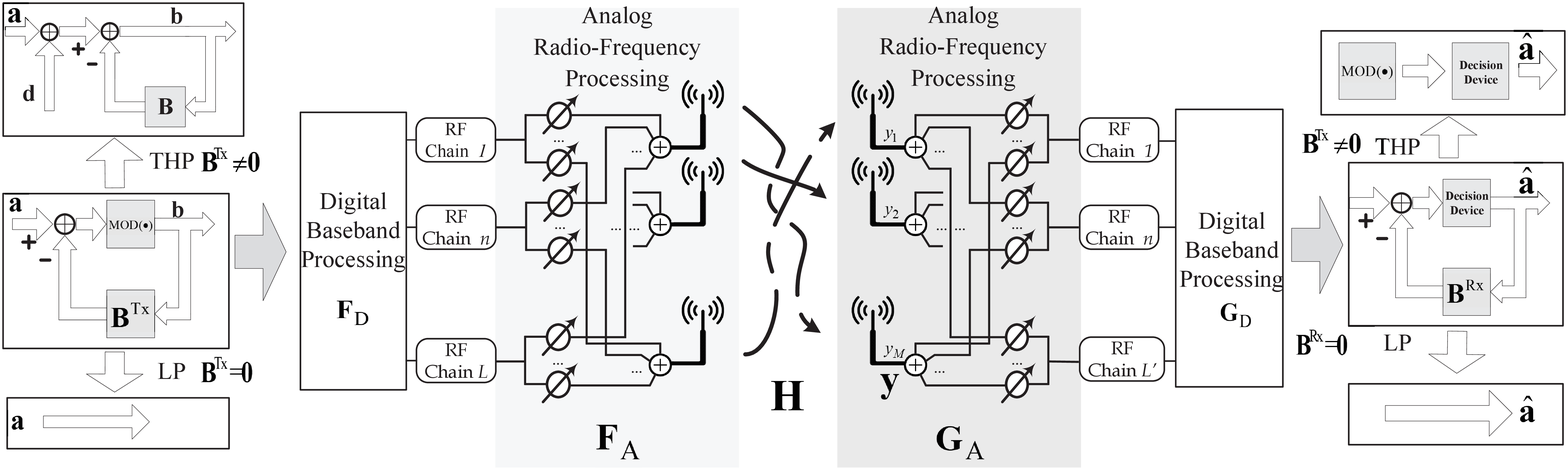}
	\caption{General hybrid MIMO transceiver.}
	\label{fig01}
\end{figure*}conclusions are drawn in Section VII.

\noindent {\textbf{Notations}}: In this paper,  scalars, vectors, and matrices are denoted by non-bold, bold lower-case, and bold upper-case letters, respectively.
The notations ${\bf X}^{\rm{H}}$ and ${\rm{Tr}}({\bf X})$ denote the Hermitian and the trace of a complex matrix ${\bf X}$, respectively.
Matrix ${\bf X}^{\frac{1}{2}}$ is the Hermitian square root of a positive semi-definite matrix ${\bf X}$.
The expression $ \mathrm{diag} \lbrace \mathbf{X} \rbrace $ denotes a square diagonal matrix with the same diagonal elements as matrix $ \mathbf{X} $.
The $ i $th row and the $ j $th column of a matrix are denoted as $ [\cdot ]_{i,:} $ and $ [\cdot ]_{:,j} $, respectively, and the element in the $ k $th row and the $ \ell $th column of a matrix is denoted as $ [\cdot ]_{k,\ell} $.
In the following derivations, ${\boldsymbol \Lambda}$ always denotes a diagonal matrix (square or rectangular diagonal matrix) with diagonal elements arranged in a nonincreasing order.
Representation $ \mathbf{A} \preceq \mathbf{B} $ means that the matrix $ \mathbf{B} - \mathbf{A} $ is positive semidefinite. The real and imaginary parts of a complex variable are represented by $ \Re \lbrace \cdot \rbrace $ and $ \Im \lbrace \cdot \rbrace $, respectively, and statistical expectation is denoted by $\mathbb{E}\{\cdot\}$.

\section{General Structure of Hybrid MIMO Transceiver}

In this section, we will first introduce the system model of MIMO hybrid transceiver designs. Then a general signal model is introduced, which includes nonlinear transceiver with THP or DFD and linear transceiver as its special cases. Based on the general signal model, the general linear minimum mean-squared error (LMMSE) processor and data estimation mean-squared error (MSE) matrix are derived, which are the basis for the subsequent hybrid MIMO transceiver design.

\subsection{System Model}

As shown in Fig.~\ref{fig01}, we consider a point-to-point hybrid MIMO system where the source and the destination are equipped with $ N $ and $ M $ antennas, respectively. Without loss of generality, it is assumed that both the source and the destination have $L$ RF chains. A transmit data vector ${\bf{a}}$ $\in \mathbb{C}^{D\times 1}$ is first processed by a unit with feedback operation and then goes  through a digital precoder $ {\bf F}_{\rm D} \in \mathbb{C}^{L \times D} $ and an analog precoder $ {\bf F}_{\rm A} \in \mathbb{C}^{N \times L} $.
This is a more general model as it includes both linear precoder and nonlinear precoder as its special cases.
For the nonlinear transceiver with THP at source, the feedback matrix $ {\bf B}^{\rm Tx} $ is strictly lower triangular. The key idea behind THP is to exploit feedback operations to pre-eliminate mutual interference between different data streams. In order to control transmit signals in a predefined region, a modulo operation is introduced for the feedback operation \cite{XingJSAC2012}. Based on lattice theory, it can be proved that the modulo operation is equivalent to adding an auxiliary complex vector ${\bf{d}}$ whose element is with integer imaginary and real parts \cite{XingJSAC2012,majorizationTHP2008}. The vector ${\bf{d}}$ makes sure  ${\bf{x}}={\bf a} + {\bf d}$ in a predefined region \cite{XingJSAC2012,majorizationTHP2008}.
Based on this fact, the output vector ${\bf{b}}$ of the feedback unit satisfies the following equation
\begin{align}
	{\bf b} = ({\bf a} + {\bf d})-{\bf{B}}^{\rm{Tx}}{\bf{b}},
\end{align}that is
\begin{align}
\label{signal_model_b}
{\bf{b}}=({\bf{I}}+{\bf{B}}^{\rm{Tx}})^{-1}\underbrace{({\bf a} + {\bf d})}_{\triangleq {\bf{x}}}.
\end{align} It is worth noting that ${\bf{d}}$ can be perfectly removed by a modulo operation \cite{XingJSAC2012,majorizationTHP2008} and thus recovering ${\bf{x}}$ is equivalent to recovering ${\bf{a}}$. On the other hand, for linear precoder, there is no feedback operation, i.e., ${\bf{B}}^{\rm{Tx}}={\bf{0}}$ and ${\bf{d}}={\bf{0}}$ \cite{XingTSP201502}. Moreover, based on (\ref{signal_model_b}) we have ${\bf{b}}={\bf{a}}$.

Then, the received signal ${\bf{y}}$ at the destination is
\begin{align}
	{\bf{y}}={\bf{H}}{\bf{F}}_{\rm{A}}{\bf{F}}_{\rm{D}}({\bf I}+{\bf B}^{\rm Tx})^{-1}{\bf{x}}+{\bf{n}},
	\label{eq-recvSig}
\end{align}
where $ {\bf n} $ is an $ M \times 1 $ additive Gaussian noise vector with zero mean and covariance $ {\bf R}_{\rm n} $, $ {\bf H} $ is an $ M \times N $ channel matrix, and $ {\bf B}^{\rm Tx} $ is a general feedback matrix at source, which is determined by the types of precoders.
It is worth noting that $ {\bf B}^{\rm Tx} = {\bf 0} $ corresponds to linear precoder without feedback operation. As shown in Fig.~\ref{fig01}, after analog and digital processing at the destination, the recovered signal is given by
\begin{align}
	{\bf{\hat x}}^{\rm{General}}={\bf{G}}_{\rm{D}}{\bf{G}}_{\rm{A}}{\bf{y}} - {\bf B}^{\rm Rx}{\bf x},
	\label{eq-dsrSig}
\end{align}
where $ {\bf G}_{\rm A} \in \mathbb{C}^{L \times M} $ is an analog processor, $ {\bf G}_{\rm D} \in \mathbb{C}^{D \times L} $ is a digital processor, and $ {\bf B}^{\rm Rx} $ is a general feedback matrix at the destination.
Note that since the analog precoder $ {\bf F}_{\rm A} $ and analog processor $ {\bf G}_{\rm A} $ are implemented through phase shifters, they are restricted to constant-modulus matrices with constant magnitude elements.
For DFD at the receiver, the decision feedback matrix $ {\bf B}^{\rm Rx} $ in (\ref{eq-dsrSig}) is a strictly lower-triangular matrix.
For linear detection, the feedback matrix in (\ref{eq-dsrSig}) is an all-zero matrix, i.e., $ {\bf B}^{\rm Rx} = {\bf 0} $.  Based on (\ref{eq-recvSig}) and (\ref{eq-dsrSig}), the recovered signal vector can be rewritten as
\begin{align}
	\label{signal_model_general}
	{\bf{\hat x}}=  {\bf{G}}_{\rm{D}}{\bf{G}}_{\rm{A}}{\bf{H}}{\bf{F}}_{\rm{A}} {\bf{F}}_{\rm{D}}( {\bf I} + {\bf B}^{\rm Tx} )^{-1}{\bf{x}} -{\bf B}^{\rm Rx}{\bf{x}} + {\bf{G}}_{\rm{D}}{\bf{G}}_{\rm{A}}{\bf{n}}.
\end{align} This is a general signal model and includes nonlinear hybrid transceivers with THP or DFD and linear hybrid transceiver as its special cases.

More specifically, for a linear hybrid transceiver, there is no feedback, either at the source or at the destination, i.e., $ {\bf B}^{\rm Tx} = {\bf B}^{\rm Rx} = {\bf 0} $. Therefore, the recovered signal in (\ref{signal_model_general}) becomes
\begin{align}
	\label{signal_model_linear}
	{\bf{\hat x}}^{\rm Linear} = {\bf{G}}_{\rm{D}}{\bf{G}}_{\rm{A}}{\bf{H}}{\bf{F}}_{\rm{A}}
	{\bf{F}}_{\rm{D}}{\bf{x}}+{\bf{G}}_{\rm{D}}{\bf{G}}_{\rm{A}}{\bf{n}}.
\end{align}

For the nonlinear transceiver with THP at the source and linear decision at the destination, i.e., $ {\bf B}^{\rm Rx} = {\bf 0} $ \cite{XingJSAC2012,majorizationTHP2008}, the detected signal vector in (\ref{signal_model_general}) becomes
\begin{align}
	\label{signal_model_thp}
	{\bf{\hat x}}^{\rm THP} = {\bf{G}}_{\rm{D}}{\bf{G}}_{\rm{A}}{\bf{H}}{\bf{F}}_{\rm{A}}
	{\bf{F}}_{\rm{D}}({\bf{I}}+{\bf{B}}^{\rm Tx})^{-1}{\bf{x}}
	+ {\bf{G}}_{\rm{D}} {\bf{G}}_{\rm{A}} {\bf{n}}.
\end{align}

For the nonlinear transceiver with DFD at the destination and a linear precoder at the source, i.e., $ {\bf B}^{\rm Tx} = {\bf 0} $, the detected signal vector in (\ref{signal_model_general}) becomes
\begin{align}
	\label{signal_model_dfe}
	{\bf{\hat x}}^{\rm DFD} = ({\bf{G}}_{\rm{D}}{\bf{G}}_{\rm{A}}{\bf{H}}{\bf{F}}_{\rm{A}}
	{\bf{F}}_{\rm{D}}-{\bf{B}}^{\rm Rx}){\bf{x}}+{\bf{G}}_{\rm{D}}
	{\bf{G}}_{\rm{A}}{\bf{n}}.
\end{align}

\subsection{Unified MSE Matrix for Different Precoders and Processors}

Based on the general signal model in (\ref{signal_model_general}), the general MSE matrix of the recovered signal at the destination equals
\begin{align}
	\label{def_MSE_Matrix}
	& {\boldsymbol{\Phi}}_{\rm{MSE}}
	({\bf{G}}_{\rm{D}},{\bf{G}}_{\rm{A}},
{\bf{F}}_{\rm{A}},{\bf{F}}_{\rm{D}},{\bf B}^{\rm Tx},{\bf B}^{\rm Rx}) \nonumber \\
	= \, & \mathbb{E}\{({\bf{\hat x}}-{\bf{x}})({\bf{\hat x}}-{\bf{x}})^{\rm{H}}\} \nonumber \\
=\, & \mathbb{E}\{\left( {\bf{G}}_{\rm{D}}{\bf{G}}_{\rm{A}}{\bf{H}}{\bf{F}}_{\rm{A}} {\bf{F}}_{\rm{D}}({\bf B}^{\rm Rx}+{\bf{I}})^{-1}{\bf{x}}-{\bf{x}}-{\bf{G}}_{\rm{D}}{\bf{G}}_{\rm{A}}{\bf{n}}  \right)\nonumber \\
&\times \left( {\bf{G}}_{\rm{D}}{\bf{G}}_{\rm{A}}{\bf{H}}{\bf{F}}_{\rm{A}} {\bf{F}}_{\rm{D}}({\bf B}^{\rm Rx}+{\bf{I}})^{-1}{\bf{x}}-{\bf{x}}-{\bf{G}}_{\rm{D}}{\bf{G}}_{\rm{A}}{\bf{n}}  \right)^{\rm{H}}\}     \nonumber \\
	= \, & \mathbb{E} \lbrace \big( {\bf{G}}_{\rm{D}}{\bf{G}}_{\rm{A}}{\bf{H}}{\bf{F}}_{\rm{A}} {\bf{F}}_{\rm{D}} - ({\bf B}^{\rm Rx}+{\bf{I}})( {\bf I} + {\bf B}^{\rm Tx} ) \big) {\bf b} {\bf b}^{\rm H}\nonumber \\
 & \times\big( {\bf{G}}_{\rm{D}}{\bf{G}}_{\rm{A}}{\bf{H}}{\bf{F}}_{\rm{A}} {\bf{F}}_{\rm{D}} - ({\bf B}^{\rm Rx}+{\bf{I}})( {\bf I} + {\bf B}^{\rm Tx} )  \big)^{\rm H} \rbrace \nonumber \\
	& \, + {\bf{G}}_{\rm{D}}{\bf{G}}_{\rm{A}}{\bf{R}}_{\rm{n}}
	{\bf{G}}_{\rm{A}}^{\rm{H}}{\bf{G}}_{\rm{D}}^{\rm{H}},
\end{align}where the third equality is based on ${\bf{b}}=({\bf B}^{\rm Rx}+{\bf{I}})^{-1}{\bf{x}}$ given in (\ref{signal_model_b}).

Based on lattice theory, each element of ${\bf b}$ is identical and independent distributed, i.e., $\mathbb{E}\{{\bf b} {\bf b}^{\rm H}\}\propto{\bf{I}}$ \cite{majorizationTHP2008}.
Thus, for notational simplicity,  we can assume $\mathbb{E}\{{\bf b} {\bf b}^{\rm H}\}={\bf{I}}$ in the following derivations. Denote ${\bf{B}}={\bf B}^{\rm Rx}+{\bf B}^{\rm Tx}+{\bf B}^{\rm Rx}{\bf B}^{\rm Tx}$, then
\begin{align}
\label{B}
 ({\bf B}^{\rm Rx}+{\bf{I}})( {\bf I} + {\bf B}^{\rm Tx} )={\bf{I}}+{\bf{B}}.
\end{align}
It is obvious that ${\bf{B}}$ is a strictly lower-triangular matrix based on the definitions of ${\bf B}^{\rm Tx}$ and  ${\bf B}^{\rm Rx}$, which implies that using nonlinear precoding at transmitter and nonlinear detection at the receiver at the same time is equivalent to just one of two. Therefore, nonlinear precoding at the transmitter and nonlinear detection at the receiver are equivalent and only one is enough.

Direct matrix derivation  \cite{XingTSP201502} yields that the optimal $ {\bf G}_{\rm D} $ will be
\begin{align}
	{\bf{G}}_{\rm D}^{\rm opt} & = ({\bf{I}}+{\bf B}) ({\bf{G}}_{\rm{A}}{\bf{H}}
	{\bf{F}}_{\rm{A}}{\bf{F}}_{\rm{D}})^{\rm{H}}\nonumber\\
& \ \ \  \times[({\bf{G}}_{\rm{A}}{\bf{H}}
	{\bf{F}}_{\rm{A}}{\bf{F}}_{\rm{D}})({\bf{G}}_{\rm{A}}{\bf{H}}
	{\bf{F}}_{\rm{A}}{\bf{F}}_{\rm{D}})^{\rm{H}}+{\bf{G}}_{\rm{A}} {\bf{R}}_{\rm{n}}
	{\bf{G}}_{\rm{A}}^{\rm{H}}]^{-1}.
	\label{eq-MMSE-Equalizer-Nonlinear}
\end{align}That is, the general MSE matrix can be further simplified into
\begin{align}
	\label{MSE_Matrix_Nonlinear}
	& {\boldsymbol{\Phi}}_{\rm{MSE}}
	({\bf{G}}_{\rm{A}},{\bf{F}}_{\rm{A}},
	{\bf{F}}_{\rm{D}},{\bf B}) \nonumber \\
	= \, & {\boldsymbol{\Phi}}_{\rm{MSE}}
	({\bf{G}}_{\rm{D}}^{\rm{opt}},{\bf{G}}_{\rm{A}},{\bf{F}}_{\rm{A}},
	{\bf{F}}_{\rm{D}},{\bf B}^{\rm Tx},{\bf B}^{\rm Rx}) \nonumber \\
	= \, &  ( {\bf I} + {\bf B})  ({\bf I} + {\bf F}_{\rm D}^{\rm H} {\bf F}_{\rm A}^{\rm H} {\bf H}^{\rm H} {\bf G}_{\rm A}^{\rm H}
	({\bf G}_{\rm A} {\bf R}_{\rm n} {\bf G}_{\rm A}^{\rm H})^{-1} {\bf G}_{\rm A} {\bf H} {\bf F}_{\rm A}{\bf F}_{\rm D})^{-1}\nonumber \\
  &\times({\bf I} + {\bf B} )^{\rm{H}} \nonumber \\
	\preceq \, & {\boldsymbol{\Phi}}_{\rm{MSE}}
	({\bf{G}}_{\rm{D}},{\bf{G}}_{\rm{A}},{\bf{F}}_{\rm{A}},
	{\bf{F}}_{\rm{D}},{\bf B}^{\rm Tx},{\bf B}^{\rm Rx}),
\end{align}for any $ {\bf G}_{\rm D} $.

If $ {\bf B}= {\bf 0} $ in (\ref{eq-MMSE-Equalizer-Nonlinear}) and (\ref{MSE_Matrix_Nonlinear}), the results are reduced to linear transceiver. Specifically, the corresponding digital LMMSE processor for linear transceiver is given as follows
\begin{align}
	{\bf{G}}_{\rm D,L}^{\rm{opt}} = & ({\bf{G}}_{\rm{A}}{\bf{H}}
	{\bf{F}}_{\rm{A}}{\bf{F}}_{\rm{D}})^{\rm{H}}[({\bf{G}}_{\rm{A}}{\bf{H}}
	{\bf{F}}_{\rm{A}}{\bf{F}}_{\rm{D}})({\bf{G}}_{\rm{A}}{\bf{H}}
	{\bf{F}}_{\rm{A}}{\bf{F}}_{\rm{D}})^{\rm{H}}\nonumber \\
&+{\bf{G}}_{\rm{A}}{\bf{R}}_{\rm{n}}
	{\bf{G}}_{\rm{A}}^{\rm{H}}]^{-1},
	\label{eq-MMSE-Equalizer-Linear}
\end{align}
and the MSE matrix for linear transceiver is
\begin{align}
	\label{MSE_Matrix_Linear}
	& {\boldsymbol{\Phi}}_{\rm{MSE}}^{\rm{L}}
	({\bf{G}}_{\rm{A}},{\bf{F}}_{\rm{A}},
	{\bf{F}}_{\rm{D}})\nonumber \\
=&{\boldsymbol{\Phi}}_{\rm{MSE}}
	({\bf{G}}_{\rm{A}},{\bf{F}}_{\rm{A}},
	{\bf{F}}_{\rm{D}},{\bf 0})\nonumber \\
	= &\left[{\bf{I}} + {\bf{F}}_{\rm{D}}^{\rm{H}} {\bf{F}}_{\rm{A}}^{\rm{H}} {\bf{H}}^{\rm{H}} {\bf{G}}_{\rm{A}}^{\rm{H}}
	({\bf{G}}_{\rm{A}}{\bf{R}}_{\rm{n}}{\bf{G}}_{\rm{A}}^{\rm{H}})^{-1} {\bf{G}}_{\rm{A}}{\bf{H}}{\bf{F}}_{\rm{A}}{\bf{F}}_{\rm{D}}\right]^{-1}\nonumber \\
\triangleq&\left[{\bf{I}}+{\boldsymbol \Gamma}({\bf{G}}_{\rm{A}}, {\bf{F}}_{\rm{D}}, {\bf{F}}_{\rm{A}})\right]^{-1},
\end{align}where
\begin{align}
\label{matrix_SNR}
{\boldsymbol \Gamma}({\bf{G}}_{\rm{A}}, {\bf{F}}_{\rm{D}}, {\bf{F}}_{\rm{A}})={\bf{F}}_{\rm{D}}^{\rm{H}} {\bf{F}}_{\rm{A}}^{\rm{H}} {\bf{H}}^{\rm{H}} {\bf{G}}_{\rm{A}}^{\rm{H}}
	({\bf{G}}_{\rm{A}}{\bf{R}}_{\rm{n}}{\bf{G}}_{\rm{A}}^{\rm{H}})^{-1} {\bf{G}}_{\rm{A}}{\bf{H}}{\bf{F}}_{\rm{A}}{\bf{F}}_{\rm{D}}\end{align},
which is signal-to-noise ratio for single antenna case.

For the nolinear transceivers, ${\bf{B}}={\bf B}^{\rm Tx}$ for THP or ${\bf{B}}={\bf B}^{\rm Rx}$ for DFD in (\ref{B})-(\ref{MSE_Matrix_Nonlinear}).
Based on (\ref{MSE_Matrix_Nonlinear}) and  (\ref{MSE_Matrix_Linear}), the general MSE matrix for nonlinear transceivers can also be written in the following unified formula
\begin{align}
\label{MSE_general_simple}
	{\boldsymbol{\Phi}}_{\rm{MSE}}
	({\bf{G}}_{\rm{A}}, {\bf{F}}_{\rm{A}}, {\bf{F}}_{\rm{D}}, {\bf{B}})
	&= ({\bf{I}} + {\bf{B}}) {\boldsymbol{\Phi}}_{\rm{MSE}}^{\rm{L}}
	({\bf{G}}_{\rm{A}}, {\bf{F}}_{\rm{A}}, {\bf{F}}_{\rm{D}}) \nonumber \\
&\times({\bf{I}} + {\bf{B}})^{\rm H},
\end{align}which  turns into the MSE matrix in (\ref{MSE_Matrix_Nonlinear}) when ${\bf{B}}={\bf{0}}$.

In the following, we will investigate unified hybrid MIMO transceiver optimization, which is applicable to various objective functions based on on the general MSE matrix (\ref{MSE_general_simple}).

\section{The Unified Hybrid MIMO Transceiver Optimization}

Because of the multi-objective optimization nature for MIMO systems with multiple data streams, there are different kinds of objectives that reflect different design preferences \cite{Palomar03}. All can be regarded as a matrix monotonic function of the data estimation MSE matrix in (\ref{MSE_general_simple}) \cite{XingTSP201501}. A function $f(\cdot)$ is a matrix monotone increasing function if $ f({\bf X}) \ge f({\bf Y}) $ for $ {\bf X} \succeq {\bf Y} \succeq {\bf 0}$ \cite{XingTSP201501}.
To avoid case-by-case discussion, we will investigate in depth hybrid MIMO transceiver optimization with different performance metrics from a unified viewpoint, in this section.

Based on the MSE matrix in (\ref{MSE_general_simple}), the unified hybrid MIMO transceiver design can be formulated in the following form
\begin{align}
	\label{General_Transceiver}
	\min_{{\bf{G}}_{\rm{A}},{\bf{F}}_{\rm{A}},{\bf{F}}_{\rm{D}},{\bf{B}}} \  & f \left(
	{\boldsymbol{\Phi}}_{\rm{MSE}}
	({\bf{G}}_{\rm{A}},{\bf{F}}_{\rm{A}},
	{\bf{F}}_{\rm{D}},{\bf{B}})
	\right) \nonumber \\
	{\rm{s.t.}} \qquad & {\rm{Tr}}({\bf{F}}_{\rm{A}}{\bf{F}}_{\rm{D}}{\bf{F}}_{\rm{D}}^{\rm{H}}{\bf{F}}_{\rm{A}}^{\rm{H}})\le P \nonumber \\
	&  {\bf{F}}_{\rm{A}} \in \mathcal{F}, \ \ {\bf{G}}_{\rm{A}} \in \mathcal{G},
\end{align}
where $f(\cdot)$ is a matrix monotone increasing function \cite{XingTSP201501}. The sets $\mathcal{F}$ and $\mathcal{G}$ are the feasible analog precoder set and analog processor set satisfying constant-modulus constraint, and $ P $ denotes the maximum transmit power at the source.


\subsection{Specific Objective Functions}
There are many ways to choose the matrix monotone increasing function. In this subsection, we will investigate the properties of different objective functions in (\ref{MSE_general_simple}).

One group of matrix monotone increasing functions can be expressed as
\begin{align}
\label{f_1}
&f_1\left(
	{\boldsymbol{\Phi}}_{\rm{MSE}}
	({\bf{G}}_{\rm{A}},{\bf{F}}_{\rm{A}},
	{\bf{F}}_{\rm{D}},{\bf{B}})
	\right)\nonumber \\
&=f_{\rm{Schur}} \left({\bf{d}}(
	{\boldsymbol{\Phi}}_{\rm{MSE}}
	({\bf{G}}_{\rm{A}},{\bf{F}}_{\rm{A}},
	{\bf{F}}_{\rm{D}},{\bf{B}}))
	\right),
\end{align}where ${\bf{d}}(
	{\boldsymbol{\Phi}}_{\rm{MSE}}
	({\bf{G}}_{\rm{A}},{\bf{F}}_{\rm{A}},
	{\bf{F}}_{\rm{D}},{\bf{B}}))$ is a vector consisting of the diagonal elements of the matrix ${\boldsymbol{\Phi}}_{\rm{MSE}}
	({\bf{G}}_{\rm{A}},{\bf{F}}_{\rm{A}},
	{\bf{F}}_{\rm{D}},{\bf{B}})$ and $f_{\rm{Schur}}(\cdot)$ is a function of a vector satisfying one the following four properties discussed in Appendix~\ref{Appendix_A}:
\begin{enumerate}
\item Multiplicatively Schur-convex
\item Multiplicatively Schur-concave
\item Additively Schur-convex
\item Additively Schur-concave.
\end{enumerate}
Many widely used metrics can be regarded as a special case of this group of functions \cite{Palomar03,majorizationTHP2008,XingJSAC2012}.

\noindent \textbf{Conclusion 1:} For linear transceiver, the feedback matrix ${\bf{B}}$ in (\ref{General_Transceiver}) is an all-zero matrix, i.e., $ {\bf B}^{\rm{opt}} = {\bf 0} $.
For nonlinear transceiver, from Appendix~\ref{Appendix_Optimal_B}  the optimal feedback matrix ${\bf{B}}$ for $f_1(\cdot)$ is
\begin{align}
	\label{optimal_B}
	{\bf{B}}^{\rm{opt}}
	={\rm{diag}}\{[[{\bf{L}}]_{1,1},\cdots,[{\bf{L}}]_{L,L}]^{\rm{T}}\}
	{\bf{L}}^{-1}-{\bf{I}},
\end{align}
where ${\bf{L}}$ is a lower triangular matrix of the following Cholesky decomposition
\begin{align}
\label{Definition_L}
	{\boldsymbol{\Phi}}_{\rm{MSE}}^{\rm{L}}
	({\bf{G}}_{\rm{A}},{\bf{F}}_{\rm{A}},{\bf{F}}_{\rm{D}})
	={\bf{L}}{\bf{L}}^{\rm{H}}.
\end{align}



%
It has been proved in \cite{XingTSP201501} and \cite{XingTSP201502} that for nonlinear transceiver design each data stream will have the same performance if $f_{\rm{Schur}}(\cdot)$ in (\ref{f_1}) is multiplicatively Schur-convex. On the other hand, if $f_{\rm{Schur}}(\cdot)$ in (\ref{f_1}) is multiplicatively Schur-concave, for nonlinear transceiver design the objective function includes geometrically weighted signal-to-noise-plus-interference-ratio (SINR) maximization as its special case.

If $f_{\rm{Schur}}(\cdot)$ in (\ref{f_1}) is additively Schur-convex, the objective function includes the the maximum MSE minimization and the minimum BER with the same constellation on each data stream as special cases. If $f_{\rm{Schur}}(\cdot)$ in (\ref{f_1}) is additively Schur-concave, the objective function includes weighted MSE minimization as its special case. Additive Schur functions are usually used for linear transceivers (${\bf{B}}={\bf{0}}$ in (\ref{f_1})) since closed-form solutions can be obtained in this case.

Besides the above group of matrix monotone increasing functions, we can choose one to reflect capacity and MSE for linear transceivers.
Capacity is one of the most popular performance metrics in MIMO transceiver optimization. It can be expressed as the form of MSE matrix considering the well-known relationship between the MSE matrix and capacity \cite{XingTSP201501}, i.e., $ C = -{\rm log}| {\bf \Phi}_{\rm MSE} | $. Then, the objective can be given as
\begin{align}
	& f_{\rm 2} (\cdot)
	= {\rm{log}}|{\boldsymbol{\Phi}}_{\rm{MSE}}^{\rm{L}}
	({\bf{G}}_{\rm{A}},{\bf{F}}_{\rm{A}},{\bf{F}}_{\rm{D}})|.
	\label{obj-cap}
\end{align}MSE is another widely used performance metric that demonstrates how accurately a signal can be recovered. The corresponding weighted MSE minimization objective is
\begin{align}
	&  f_{\rm 3} (\cdot)
	= {\rm{Tr}}\left[{\bf{A}}^{\rm{H}}
	{\boldsymbol{\Phi}}_{\rm{MSE}}^{\rm{L}}
	({\bf{G}}_{\rm{A}},{\bf{F}}_{\rm{A}},{\bf{F}}_{\rm{D}}){\bf{A}}\right],
	\label{obj-wmse}
\end{align}
where $ {\bf A} $ is a general, not necessarily diagonal, weight matrix, even if it is often diagonal in many applications.


\subsection{Hybrid MIMO Transceiver Optimization}

Denote
\begin{align}
	{\boldsymbol{\Pi}}_{\rm{L}} & =
	( {\bf{G}}_{\rm{A}} {\bf{R}}_{\rm{n}} {\bf{G}}_{\rm{A}}^{\rm{H}} )^{-1/2}{\bf{G}}_{\rm{A}}{\bf{R}}_{\rm{n}}^{1/2}, \nonumber \\
	{\boldsymbol{\Pi}}_{\rm{R}} & = {\bf{F}}_{\rm{A}} ( {\bf{F}}_{\rm{A}}^{\rm{H}} {\bf{F}}_{\rm{A}} )^{-\frac{1}{2}}, \nonumber
\end{align}
and
\begin{align}
	{\bf{\tilde F}}_{\rm{D}} & = ( {\bf{F}}_{\rm{A}}^{\rm{H}} {\bf{F}}_{\rm{A}} )^{\frac{1}{2}}{\bf{ F}}_{\rm{D}}{\bf{Q}}^{\rm{H}},
	\label{def-PI}
\end{align}
where ${\bf{Q}}$ is a unitary matrix to be determined by digital transceiver optimization in the next section. Then (\ref{matrix_SNR}) can be rewritten as
\begin{align}	{\boldsymbol{\Gamma}}({\bf{G}}_{\rm{A}},{\bf{F}}_{\rm{D}},{\bf{F}}_{\rm{A}})
	={\bf{Q}}^{\rm{H}}{\boldsymbol{\tilde \Gamma}}({\bf{G}}_{\rm{A}},{\bf{\tilde F}}_{\rm{D}},{\bf{F}}_{\rm{A}}){\bf{Q}},
	\label{eq-SNR}
\end{align}
where
\begin{align}
	& {\boldsymbol{\tilde \Gamma}}({\bf{G}}_{\rm{A}},{\bf{\tilde F}}_{\rm{D}},{\bf{F}}_{\rm{A}})
	=  {\bf{\tilde F}}_{\rm{D}}^{\rm{H}}{\boldsymbol{\Pi}}_{\rm{R}}^{\rm{H}}
	{\bf{H}}^{\rm{H}}{\bf{R}}_{\rm{n}}^{-1/2}
	{\boldsymbol{\Pi}}_{\rm{L}}^{\rm{H}}
	{\boldsymbol{\Pi}}_{\rm{L}} {\bf{R}}_{\rm{n}}^{-1/2}{\bf{H}}{\boldsymbol{\Pi}}_{\rm{R}}{\bf{\tilde F}}_{\rm{D}}.
\end{align}

The optimal ${\bf{B}}$ is usually a function of ${\boldsymbol{\Gamma}}({\bf{G}}_{\rm{A}},
{\bf{F}}_{\rm{D}},{\bf{F}}_{\rm{A}})$, for all objective functions as demonstrated by (\ref{optimal_B}) for $f_{\rm{Schur}}(\cdot)$ in (\ref{f_1}). From (\ref{eq-SNR}), we can conclude that  the optimal ${\bf{B}}$ is a function of ${\bf{Q}}^{\rm{H}}{\boldsymbol{\tilde \Gamma}}({\bf{G}}_{\rm{A}},{\bf{\tilde F}}_{\rm{D}},{\bf{F}}_{\rm{A}}){\bf{Q}}$.
Therefore, using (\ref{MSE_Matrix_Linear}) and (\ref{eq-SNR}), the objective function of (\ref{General_Transceiver}) can be expressed in terms of ${\boldsymbol{\tilde \Gamma}}({\bf{G}}_{\rm{A}},{\bf{\tilde F}}_{\rm{D}},{\bf{F}}_{\rm{A}})$ as
\begin{align}
\label{function_function}
	& f \left( {\boldsymbol{\Phi}}_{\rm{MSE}}
	({\bf{G}}_{\rm{A}},{\bf{F}}_{\rm{A}},{\bf{F}}_{\rm{D}},{\bf{B}}^{\rm{opt}}) \right) \nonumber \\
	= & f \big( ({\bf{I}}+{\bf{B}}^{\rm opt}) ({\bf I} + {\bf{Q}}^{\rm{H}}{\boldsymbol{\tilde \Gamma}} ({\bf{G}}_{\rm{A}},{\bf{\tilde F}}_{\rm{D}},{\bf{F}}_{\rm{A}}) {\bf{Q}} )^{-1} ({\bf{I}}+{\bf{B}}^{\rm opt})^{\rm{H}} \big) \nonumber \\
	\triangleq & f_{{\rm S}} \left({\bf{Q}}^{\rm{H}}{\boldsymbol{\tilde \Gamma}}({\bf{G}}_{\rm{A}},{\bf{\tilde F}}_{\rm{D}},{\bf{F}}_{\rm{A}}){\bf{Q}}\right).
\end{align}
After introducing ${\boldsymbol{\tilde \Gamma}}({\bf{G}}_{\rm{A}},{\bf{\tilde F}}_{\rm{D}},{\bf{F}}_{\rm{A}})$ and a new auxiliary matrix $ {\bf Q} $, the objective function is transferred into $ f_{{\rm S}}(\cdot)$ rather than $ f(\cdot) $. Note that this new function notation, $ f_{{\rm S}}(\cdot)$, is defined only for notational simplicity and it explicitly expresses the objective as a function of matrix variables ${\bf Q}$ and ${\boldsymbol{\tilde \Gamma}}$.
Therefore, the optimization problem in (\ref{General_Transceiver}) is further rewritten into the following one
\begin{align}
	\label{unified_transceiver}
	\min_{{\bf{Q}},{\bf{G}}_{\rm{A}},{\bf{\tilde F}}_{\rm{D}},{\bf{F}}_{\rm{A}}} \ & f_{{\rm S}} \left({\bf{Q}}^{\rm{H}}{\boldsymbol{\tilde \Gamma}}({\bf{G}}_{\rm{A}},{\bf{\tilde F}}_{\rm{D}},{\bf{F}}_{\rm{A}}){\bf{Q}}\right) \nonumber \\
	{\rm{s.t.}}  \ \ \ \ \ \  & {\rm{Tr}}({\bf{\tilde F}}_{\rm{D}}{\bf{\tilde F}}_{\rm{D}}^{\rm{H}})\le P \nonumber \\
	& {\bf{F}}_{\rm{A}} \in \mathcal{F}, \ \ {\bf{G}}_{\rm{A}} \in \mathcal{G}.
\end{align} We will discuss in detail how to solve the optimization problem (\ref{unified_transceiver}) with respect to ${\bf{Q}}$, ${\bf{G}}_{\rm{A}}$,${\bf{\tilde F}}_{\rm{D}}$, and ${\bf{F}}_{\rm{A}}$ subsequently. In (\ref{unified_transceiver}), ${\bf{B}}$ has been formulated as a function of  ${\bf{Q}}$, ${\bf{G}}_{\rm{A}}$,${\bf{\tilde F}}_{\rm{D}}$, and ${\bf{F}}_{\rm{A}}$. When ${\bf{Q}}$, ${\bf{G}}_{\rm{A}}$,${\bf{\tilde F}}_{\rm{D}}$, and ${\bf{F}}_{\rm{A}}$ are calculated, the optimal ${\bf{B}}$ can be directly derived based on (\ref{optimal_B}).

\section{Digital Transceiver Optimization}

In the following, we focus on the digital transceiver optimization for the optimization problem (\ref{unified_transceiver}). More specifically, we first derive the optimal unitary matrix ${\bf{Q}}$ and then find the optimal ${\bf{\tilde F}}_{\rm{D}}$.

\subsection{Optimal $ {\bf{Q}} $}
At the beginning of this section, two fundamental definitions are given based on the following eigenvalue decomposition (EVD) and SVD
\begin{align}
	\label{EVD_SVD}
	&{\boldsymbol{\tilde \Gamma}}({\bf{G}}_{\rm{A}},{\bf{\tilde F}}_{\rm{D}},{\bf{F}}_{\rm{A}})={\bf{U}}_{{\boldsymbol{\tilde \Gamma}}}{\boldsymbol \Lambda}_{{\boldsymbol{\tilde \Gamma}}}{\bf{U}}_{{\boldsymbol{\tilde \Gamma}}}^{\rm{H}} \nonumber  \\
	&{\bf{A}}={\bf{U}}_{{\bf{A}}}{\boldsymbol \Lambda}_{{\bf{A}}}{\bf{V}}_{{\bf{A}}}^{\rm{H}},
\end{align}
where $ {\boldsymbol{\Lambda}}_{\rm \Phi} $ and $ {\boldsymbol{\Lambda}}_{\rm A} $ denote a diagonal matrix with the diagonal elements in nondecreasing order.

Denote ${\bf{U}}_{\rm{GMD}}$ as the unitary matrix that makes the lower triangular matrix ${\bf{L}}$ in (\ref{Definition_L}) has the same diagonal elements. It has been shown in \cite{Palomar03,XingTSP201501,XingTSP201502} that the optimal ${\bf{Q}}$ for the first group of matrix-monotonic functions can be expressed as
\begin{align}
\label{optimal_Q_1}
{\bf{Q}}^{\rm{opt}} = \left\{ {\begin{array}{l}
{{\bf{U}}_{{\boldsymbol{\tilde \Gamma}}}{\bf{U}}_{\text{GMD}}^{\rm{H}}}
 \ \ \text{if} \ f(\cdot) \ \text{is multiplicatively Schur-convex} \\
{{\bf{U}}_{{\boldsymbol{\tilde \Gamma}}}} \ \ \ \ \ \ \ \ \ \text{if} \ f(\cdot) \ \text{is multiplicatively Schur-concave} \\
{{\bf{U}}_{{\boldsymbol{\tilde \Gamma}}}{\bf{U}}_{\text{DFT}}^{\rm{H}}}
\ \ \ \text{if} \ f(\cdot) \ \text{is additively Schur-convex}
\\
{{\bf{U}}_{{\boldsymbol{\tilde \Gamma}}}}  \ \ \ \ \ \ \ \ \ \text{if} \ f(\cdot) \ \text{is additively Schur-concave}.
\end{array}} \right.
\end{align}
The above results are obtained by directly manipulating with the objective function $f(\cdot)$ in \eqref{function_function}, and thus the optimal ${\bf{Q}}$ varies with the matrix-monotone increasing function in (\ref{General_Transceiver}).

For the capacity maximization in (\ref{obj-cap}),  the objective function of (\ref{unified_transceiver}) can be written as
\begin{align}
	\label{Q_X_Obj_1}
	& f_{\rm S, 2}(\cdot) = -{\rm{log}} |{\bf{Q}}^{\rm{H}} {\boldsymbol{\tilde \Gamma}} ({\bf{G}}_{\rm{A}},{\bf{\tilde F}}_{\rm{D}},{\bf{F}}_{\rm{A}}) {\bf{Q}} + {\bf{I}}|.
\end{align}
Since the function in (\ref{Q_X_Obj_1}) is independent of $ {\bf Q} $ as long as it is a unitary matrix, the optimal $ {\bf Q} $, namely $ {\bf Q}^{\rm opt} $, can be any unitary matrix with proper dimension.

For the weighted MSE minimization given by (\ref{obj-wmse}), the objective function of (\ref{unified_transceiver}) can be rewritten as
\begin{align}
	f_{\rm S, 3} (\cdot) & =
	{\rm{Tr}} [{\bf{A}}^{\rm{H}} ({\bf{Q}}^{\rm{H}} {\boldsymbol{\tilde \Gamma}} ({\bf{G}}_{\rm{A}},{\bf{\tilde F}}_{\rm{D}},{\bf{F}}_{\rm{A}}) {\bf{Q}}
	+ {\bf{I}})^{-1} {\bf{A}}].
\end{align}
Based on the EVD and SVD defined in (\ref{EVD_SVD}) and the matrix inequality in Appendix~\ref{Appendix_Inequality}, the optimal ${\bf{Q}}$ is
\begin{align}
\label{optimal_Q_3}
{\bf{Q}}^{\rm opt} = {\bf{U}}_{{\boldsymbol{\tilde \Gamma}}}{\bf{ U}}_{{\bf{A}}}^{\rm{H}}.
\end{align}

We have to stress that it is still hard to find the closed-form expression for the optimal ${\bf{Q}}$ for an arbitrary function $f(\cdot)$. However, most of the meaningful and popular metric functions have been shown included in one of the above function families, and are with the closed-form expression for optimal ${\bf{Q}}$.

\subsection{Optimal ${\bf{\tilde  F}}_{\rm{D}}$}

After substituting the optimal ${\bf{Q}}$ into the objective function of (\ref{unified_transceiver}),
the objective function becomes a function of the eigenvalues of  ${\boldsymbol{\tilde \Gamma}}({\bf{G}}_{\rm{A}}, {\bf{\tilde F}}_{\rm{D}}, {\bf{F}}_{\rm{A}})$, i.e.,
\begin{align}
	& f_{\rm{S}}\left({\bf{Q}}^{\rm{H}}{\boldsymbol{\tilde \Gamma}}({\bf{G}}_{\rm{A}},{\bf{\tilde F}}_{\rm{D}},{\bf{F}}_{\rm{A}}){\bf{Q}}\right)
	\triangleq f_{\rm{E}} \left({\boldsymbol\lambda} \Big( {\boldsymbol{\tilde \Gamma}}({\bf{G}}_{\rm{A}}, {\bf{\tilde F}}_{\rm{D}}, {\bf{F}}_{\rm{A}}) \Big) \right),
	\label{obj-Opt-Fd}
\end{align}
where ${\boldsymbol \lambda}({\boldsymbol{X}})=[\lambda_1({\boldsymbol{X}}), \cdots, \lambda_L({\boldsymbol{X}})]^{\rm{T}}$ and $\lambda_i({\boldsymbol{X}})$ is the $ i $th largest eigenvalue of ${\boldsymbol{X}}$. It is worth highlighting that for $f_{{\rm{S}},1}(\cdot)$ and $f_{{\rm{S}},3}(\cdot)$ based on (\ref{optimal_Q_1}) and (\ref{optimal_Q_3}) we can directly have (\ref{obj-Opt-Fd}). For $f_{{\rm{S}},2}(\cdot)$
the optimal ${\bf{Q}}$ can be an arbitrary unitary matrix, minimizing $f_{{\rm{S}},2}(\cdot)$ mathematically equals to minimizing $-\sum_{l=1}^L{\rm{log}}\left(1+{\lambda}_l ( {\boldsymbol{\tilde \Gamma}}({\bf{G}}_{\rm{A}}, {\bf{\tilde F}}_{\rm{D}}, {\bf{F}}_{\rm{A}}))\right)$ for any ${\bf{Q}}$. In other words, (\ref{obj-Opt-Fd}) always holds for these kinds of functions discussed above.

Note that the definition in (\ref{obj-Opt-Fd}) follows from the facts that the unitary matrix in ${\boldsymbol{\tilde \Gamma}}({\bf{G}}_{\rm{A}}, {\bf{\tilde F}}_{\rm{D}}, {\bf{F}}_{\rm{A}})$ has been removed by the optimal ${\bf{Q}}$ and only its eigenvalues remain to be optimized.
Therefore, the unified hybrid MIMO transceiver optimization in (\ref{unified_transceiver}) is simplified to
\begin{align}
	\label{unified_transceiver_eigenvalue}
	\min_{{\bf{G}}_{\rm{A}},{\bf{\tilde F}}_{\rm{D}},{\bf{F}}_{\rm{A}}} \ \ & f_{\rm{E}}\left({\boldsymbol\lambda}\left({\boldsymbol{\tilde \Gamma}}({\bf{G}}_{\rm{A}},{\bf{\tilde F}}_{\rm{D}},{\bf{F}}_{\rm{A}})\right)\right) \nonumber \\
	{\rm{s.t.}}  \ \ \ & {\rm{Tr}}({\bf{\tilde F}}_{\rm{D}}{\bf{\tilde F}}_{\rm{D}}^{\rm{H}})\le P \nonumber \\
	& {\bf{F}}_{\rm{A}} \in \mathcal{F}, \ \ {\bf{G}}_{\rm{A}} \in \mathcal{G}.
\end{align}
By applying the obtained results of ${\bf Q}^{\rm opt}$ and the fact that $ f(\cdot) $ is a matrix-monotone increasing function, it can be concluded from the discussion in \cite{Palomar03,XingTSP201502} that $ f_{{\rm E}}(\cdot) $ is a vector-decreasing function for $f_{{\rm{S}},1}(\cdot)$. Moreover, substituting the optimal ${\bf{Q}}$ into the objective function of (\ref{unified_transceiver}), for
$f_{{\rm{S}},2}(\cdot)$ and $f_{{\rm{S}},3}(\cdot)$ we have
\begin{align}
f_{{\rm{E}}}(\cdot)&=-\sum_{l=1}^L{\rm{log}}\left(1+{\lambda}_l ( {\boldsymbol{\tilde \Gamma}}({\bf{G}}_{\rm{A}}, {\bf{\tilde F}}_{\rm{D}},
{\bf{F}}_{\rm{A}}))\right), \\
f_{{\rm{E}}}(\cdot)&=\sum_{l=1}^L\frac{\lambda_l({\bf{A}})}{1+{\lambda}_l ( {\boldsymbol{\tilde \Gamma}}({\bf{G}}_{\rm{A}}, {\bf{\tilde F}}_{\rm{D}}, {\bf{F}}_{\rm{A}}))},
\end{align}respectively, which implies that $ f_{{\rm E}}(\cdot) $ is also vector-decreasing. In a nutshell, based on ${\bf Q}^{\rm opt}$ we can conclude that $ f_{{\rm E}}(\cdot) $  in (\ref{unified_transceiver_eigenvalue}) is a vector-decreasing function. Thus, from (\ref{unified_transceiver_eigenvalue}), the optimization becomes maximizing the eigenvalues of ${\boldsymbol{\tilde \Gamma}}({\bf{G}}_{\rm{A}},{\bf{\tilde F}}_{\rm{D}},{\bf{F}}_{\rm{A}})$. Each eigenvalue of ${\boldsymbol{\tilde \Gamma}}({\bf{G}}_{\rm{A}},{\bf{\tilde F}}_{\rm{D}},{\bf{F}}_{\rm{A}})$ corresponds to SNR of an eigenchannel.

In problem (\ref{unified_transceiver_eigenvalue}), the variables are still matrix variables. To simplify the optimization, we will first derive the diagonalizable structure of the optimal matrix variables. Based on the derived optimal structure, the dimensionality of the optimization problem are reduced significantly.

In order to derive the optimal structure and to avoid tedious case-by-case discussion, we consider a multi-objective optimization problem in the following. Its Pareto optimal solution set contains all the optimal solutions of different types of transceiver optimizations.
In particular, as discussed in \cite{XingTSP201501}, the optimal solution of problem  (\ref{unified_transceiver_eigenvalue}) with a specific objective function, i.e., $f_1(\cdot)$, $f_{2}(\cdot)$, or $f_3(\cdot)$,  must be in the Pareto optimal solution set of the following vector optimization (multi-objective) problem
\begin{align}
	\label{MM_Eigen}
	& \ \ \max_{{\bf{G}}_{\rm{A}},{\bf{\tilde F}}_{\rm{D}},{\bf{F}}_{\rm{A}}} \ \ {\boldsymbol\lambda}\left({\boldsymbol{\tilde \Gamma}}({\bf{G}}_{\rm{A}},{\bf{\tilde F}}_{\rm{D}},{\bf{F}}_{\rm{A}})\right) \nonumber \\
	& \ \ \ \ \ \ \  {\rm{s.t.}} \ \ \ \ \ \ {\rm{Tr}}({\bf{\tilde F}}_{\rm{D}}{\bf{\tilde F}}_{\rm{D}}^{\rm{H}})\le P \nonumber \\
	& \ \ \ \ \ \ \ \ \ \ \ \ \ \ \ \ \ {\bf{F}}_{\rm{A}} \in \mathcal{F}, \ \ {\bf{G}}_{\rm{A}} \in \mathcal{G}.
\end{align}Equivalently,
the vector optimization problem in (\ref{MM_Eigen}) can be rewritten as the following matrix-monotonic optimization problem
\begin{align}
	\label{MM_SNR}
	& \ \ \max_{{\bf{G}}_{\rm{A}},{\bf{\tilde F}}_{\rm{D}},{\bf{F}}_{\rm{A}}} \ \ {\boldsymbol{\tilde \Gamma}}({\bf{G}}_{\rm{A}},{\bf{\tilde F}}_{\rm{D}},{\bf{F}}_{\rm{A}}) \nonumber \\
	& \ \ \ \ \ \ \  {\rm{s.t.}} \ \ \ \ \ {\rm{Tr}}({\bf{\tilde F}}_{\rm{D}}{\bf{\tilde F}}_{\rm{D}}^{\rm{H}})\le P \nonumber \\
	& \ \ \ \ \ \ \ \ \ \ \ \ \ \ \ \ {\bf{F}}_{\rm{A}} \in \mathcal{F}, \ \  {\bf{G}}_{\rm{A}} \in \mathcal{G}.
\end{align}
It is worth noting that optimization (\ref{MM_SNR}) aims at maximizing a positive semi-definite matrix. Generally speaking, maximizing a positive semi-definite matrix includes two tasks, i.e., maximizing its eigenvalues and choosing a proper EVD unitary matrix. Note that in (\ref{MM_SNR}) there is no need to optimize the EVD unitary matrix, because the constraints can remain satisfied if only EVD unitary matrix changes. Using the definitions in (\ref{def-PI}) and given analog precoder ${\bf{F}}_{\rm{A}}$ and analog processor ${\bf{G}}_{\rm{A}}$, problem (\ref{MM_SNR}) is a standard matrix-monotonic optimization with respect to ${\bf{\tilde F}}_{\rm{D}}$. It follows
\begin{align}\label{optProblemF}
	& \ \ \max_{{\bf{\tilde F}}_{\rm{D}}} \ \ {\bf{\tilde F}}_{\rm{D}}^{\rm{H}}{\boldsymbol{\Pi}}_{\rm{R}}^{\rm{H}}{\bf{H}}^{\rm{H}}{\bf{R}}_{\rm{n}}^{-1/2}
	{\boldsymbol{\Pi}}_{\rm{L}}^{\rm{H}}
	{\boldsymbol{\Pi}}_{\rm{L}} {\bf{R}}_{\rm{n}}^{-1/2}{\bf{H}}{\boldsymbol{\Pi}}_{\rm{R}}{\bf{\tilde F}}_{\rm{D}} \nonumber \\
	&  \ \ \  {\rm{s.t.}} \ \ \ \  {\rm{Tr}}({\bf{\tilde F}}_{\rm{D}}{\bf{\tilde F}}_{\rm{D}}^{\rm{H}})\le P.
\end{align}
Based on the matrix-monotonic optimization theory developed in \cite{XingTSP201501}, the optimal solution of \eqref{optProblemF} satisfies the following diagonalizable structure.

\noindent \textbf{Conclusion 2:} Defining the following SVD,
\begin{align}
	&{\boldsymbol{\Pi}}_{\rm{L}}{\bf{R}}_{\rm{n}}^{-1/2}{\bf{H}}{\boldsymbol{\Pi}}_{\rm{R}}
	=
	{\bf{U}}_{{\boldsymbol{\mathbb{H}}}}
	{\boldsymbol \Lambda}_{\boldsymbol{\mathbb{H}}}{\bf{V}}_{\boldsymbol{\mathbb{H}}}^{\rm{H}},
\end{align} with the diagonal elements of ${\boldsymbol \Lambda}_{\boldsymbol{\mathbb{H}}}$ in decreasing order,
the optimal ${\bf{\tilde F}}_{\rm{D}}$ satisfies
\begin{align}
	{\bf{\tilde F}}_{\rm{D}}^{\rm opt} = {\bf{V}}_{\boldsymbol{\mathbb{H}}} {\boldsymbol \Lambda}_{\bf{F}} {\bf{U}}_{\rm{Arb}}^{\rm{H}},
\end{align}
where ${\boldsymbol \Lambda}_{\bf{F}}$ is a diagonal matrix determined by the specific objective functions, e.g., sum MSE, capacity maximization, etc., as discussed in the previous section.
The unitary matrix ${\bf{U}}_{\rm{Arb}}$ can be an arbitrary unitary matrix.

Thus far by using Conclusion 2, the optimal ${\bf{\tilde F}}_{\rm{D}}$ can be obtained by conducting basic manipulations as in \cite{XingTSP201501} on optimizing ${\boldsymbol \Lambda}_{\bf{F}}$ given a specific objective function.
As a result, the remaining  key task is to optimize the analog precoder and processor, which is the focus of the following section.

\section{Analog Transceiver Optimization}

Based on the optimal solution of digital precoder given in the previous section, we optimize the analog precoder and processor under constant-modulus constraints. In the following, the optimal structure of the analog transceiver is first derived. Different from existing works, we show that the analog precoder and processor design can be decoupled by using the optimal transceiver structure. This optimal structure greatly simplifies the involved analog transceiver design.

For the analog transceiver optimization in (\ref{MM_SNR}) and using (\ref{def-PI}), we have the following matrix-monotonic optimization problem
\begin{align}
	& \ \ \max_{{\bf{F}}_{\rm{A}},{\bf{G}}_{\rm{A}}} \ \ {\bf{\tilde F}}_{\rm{D}}^{\rm{H}}{\boldsymbol{\Pi}}_{\rm{R}}^{\rm{H}}{\bf{H}}^{\rm{H}}{\bf{R}}_{\rm{n}}^{-1/2}
	{\boldsymbol{\Pi}}_{\rm{L}}^{\rm{H}}
	{\boldsymbol{\Pi}}_{\rm{L}} {\bf{R}}_{\rm{n}}^{-1/2}{\bf{H}}{\boldsymbol{\Pi}}_{\rm{R}}{\bf{\tilde F}}_{\rm{D}} \nonumber \\
	&  \ \ \  {\rm{s.t.}} \ \ \ \ \ \ {\bf{F}}_{\rm{A}} \in \mathcal{F}, \ \ {\bf{G}}_{\rm{A}} \in \mathcal{G}.
	\label{eq-analog-problem}
\end{align}

Denote the SVDs
\begin{align}
	\label{effective_channel} {\bf{R}}_{\rm{n}}^{-1/2}{\bf{H}}& \triangleq {\bf{U}}_{{\boldsymbol{\mathcal{H}}}}
	{\boldsymbol \Lambda}_{\boldsymbol{\mathcal{H}}}{\bf{V}}_{\boldsymbol{\mathcal{H}}}^{\rm{H}},\\
{\bf{R}}_{\rm{n}}^{1/2} {\bf{G}}_{\rm{A}}^{\rm{H}}& \triangleq {\bf{U}}_{{\bf{R}}{\bf{G}}} {\boldsymbol{\Lambda}}_{{\bf{R}}{\bf{G}}}{\bf{V}}_{{\bf{R}}{\bf{G}}}^{\rm{H}}.
\end{align} In Appendix~\ref{Appendix_Analog_Transceiver}, we prove the following conclusion on the optimal structure of ${\bf{F}}_{\rm{A}}$ and ${\bf{G}}_{\rm{A}}$.

\noindent \textbf{Conclusion 3:} Let the SVD of ${\bf{F}}_{\rm{A}}$ be
\begin{align}
	{\bf{F}}_{\rm{A}} \triangleq {\bf{U}}_{{\bf{F}}_{\rm{A}}}
	{\boldsymbol{\Lambda}}_{{\bf{F}}_{\rm{A}}}{\bf{V}}_{{\bf{F}}_{\rm{A}}}^{\rm{H}}.
\end{align}
The singular values in ${\boldsymbol{\Lambda}}_{{\bf{F}}_{\rm{A}}}$ do not affect the objective function in \eqref{eq-analog-problem}, and the unitary matrix ${\bf{U}}_{{\bf{F}}_{\rm{A}}}$ for the optimal ${\bf{F}}_{\rm{A}}$ satisfies
\begin{align}
\label{Analog_Precoder_Optimization} [{\bf{U}}_{{\bf{F}}_{\rm{A}}}]_{:,1:L}^{\rm{opt}}={\rm{arg\,max}}
	\{
	\| [{\bf{V}}_{\boldsymbol{\mathcal{H}}}]_{:,1:L} [{\bf{U}}_{{\bf{F}}_{\rm{A}}}]_{:,1:L}^{\rm{H}}\|_{\rm{F}}^2
	\}.
\end{align}
On the other hand, denote the SVD of $ {\bf{R}}_{\rm{n}}^{1/2} {\bf{G}}_{\rm{A}}^{\rm{H}} $ as
\begin{align}
	{\bf{R}}_{\rm{n}}^{1/2} {\bf{G}}_{\rm{A}}^{\rm{H}} \triangleq {\bf{U}}_{{\bf{R}}{\bf{G}}} {\boldsymbol{\Lambda}}_{{\bf{R}}{\bf{G}}}{\bf{V}}_{{\bf{R}}{\bf{G}}}^{\rm{H}}.
\end{align}
The singular values in ${\boldsymbol{\Lambda}}_{{\bf{R}}{\bf{G}}}$ do not affect the objective in \eqref{eq-analog-problem}, and the unitary matrix ${\bf{U}}_{{\bf{R}}{\bf{G}}}$ for the optimal ${\bf{G}}_{\rm{A}}$ satisfies
\begin{align}
	[{\bf{U}}_{{\bf{R}}{\bf{G}}}]_{:,1:L}^{\rm{opt}}=
{\rm{arg\,max}}\{ \|
	[{\bf{U}}_{\boldsymbol{\mathcal{H}}}]_{:,1:L} [{\bf{U}}_{{\bf{R}}{\bf{G}}}]_{:,1:L}^{\rm{H}}\|_{\rm{F}}^2\}.
\end{align}

Based on the optimal structure given in Conclusion 3, in the following two kinds of algorithms are proposed to compute the analog precoder and processor. The first one is based on phase projection, which provides better performance while the second one based on a heuristic random selection, is with low complexity.

\subsection{Phase Projection Based Algorithm}

\noindent \textbf{Analog Precoder Design}

From Conclusion 3, the optimal analog precoder should select the first $L$-best eigenchannels. It is challenging to directly optmize ${\bf{F}}_{\rm{A}}$ based on (\ref{Analog_Precoder_Optimization}) because of the SVD of a constant-modulus matrix.
Alternatively, we resort to finding a matrix in the constant-modulus space with the minimum distance to the space spanned by $[{\bf{V}}_{\boldsymbol{\mathcal{H}}}]_{:,1:L} $. Then, the corresponding optimization problem of analog precoder design can be formulated as
\begin{align} &\min_{{\bf{F}}_{\rm{A}},{\mathrm{\mathbf{\Lambda}_A}},{\bf{Q}}_{\rm{A}}} \ \ \|[{\bf{V}}_{\boldsymbol{\mathcal{H}}}]_{:,1:L} {\mathrm{\mathbf{\Lambda}_A}}{\bf{Q}}_{\rm{A}}-{\bf{F}}_{\rm{A}}\|_{\rm{F}}^2
	\nonumber \\
	& \ \ \ \ {\rm{s.t.}} \ \ \ \ \ \  {\bf{Q}}_{\rm{A}}{\bf{Q}}_{\rm{A}}^{\rm{H}}={\bf{I}}\nonumber \\
	& \ \ \ \ \ \ \ \ \ \ \ \ \ \ {\bf{F}}_{\rm{A}} \in \mathcal{F}.
	\label{eq-pre-org}
\end{align}
Different from the existing work \cite{Y.C.Eldar2017}, the diagonal matrix $ \mathrm{\mathbf{\Lambda}_A} $ and the unitary matrix $ \mathrm{\mathbf{Q}_A} $ in our work are jointly optimized to
make $ {\bf F}_{\rm A} $ as close as possible in the space spanned by $ [{\bf{V}}_{\boldsymbol{\mathcal{H}}}]_{:,1:L} $ in terms of Frobenius norm.
As there is no constraint on the diagonal matrix $ \mathrm{\mathbf{\Lambda}_A} $, given matrices $ {\bf Q}_{\rm A} $ and $ {\bf F}_{\rm A} $, the optimal $ \mathbf{\Lambda}_{\rm A} $ is
\begin{align}
\mathbf{\Lambda}_{\rm{A}}^{\rm{opt}}= \mathrm{diag} \Big\lbrace \Re \big( {\mathrm{\mathbf{Q}_A}} {\mathrm{\mathbf{F}_A^H}} [{\bf{V}}_{\boldsymbol{\mathcal{H}}}]_{:,1:L} \big) \Big\rbrace.
	\label{eq-diagonal-matrix}
\end{align}

Then we rewrite the objective function in (\ref{eq-pre-org}) as
\begin{align}
\label{aa_MSE}
&\|[{\bf{V}}_{\boldsymbol{\mathcal{H}}}]_{:,1:L} \mathbf{\Lambda}_{\rm{A}}{\bf{Q}}_{\rm{A}}-{\bf{F}}_{\rm{A}}\|_{\rm{F}}^2 \nonumber \\
=&{\rm{Tr}}
([{\bf{V}}_{\boldsymbol{\mathcal{H}}}]_{:,1:L}
\mathbf{\Lambda}_{\rm{A}}^{\rm{opt}}(\mathbf{\Lambda}_{\rm{A}}^{\rm{opt}})
^{\rm{H}}[{\bf{V}}_{\boldsymbol{\mathcal{H}}}]_{:,1:L}^{\rm{H}})\nonumber \\
&
+{\rm{Tr}}({\bf{F}}_{\rm{A}}{\bf{F}}_{\rm{A}}^{\rm{H}})-2\Re\{{\rm{Tr}}({\bf{F}}_{\rm{A}}^{\rm{H}}[{\bf{V}}_{\boldsymbol{\mathcal{H}}}]_{:,1:L} \mathbf{\Lambda}_{\rm{A}}^{\rm{opt}}{\bf{Q}}_{\rm{A}})\}.
\end{align}To minimize (\ref{aa_MSE}) given ${\boldsymbol \Lambda}_{\rm{A}}$ and ${\bf{F}}_{\rm{A}}$, the term $\Re\{{\rm{Tr}}({\bf{F}}_{\rm{A}}^{\rm{H}}[{\bf{V}}_{\boldsymbol{\mathcal{H}}}]_{:,1:L} {\mathrm{\mathbf{\Lambda}_A}^{\rm{opt}}}{\bf{Q}}_{\rm{A}})\}$ should be maximized. By applying the matrix inequality \cite{inequalityMajorBook}, the optimal $ {\bf Q}_{\rm A} $ is
\begin{align}
	\mathbf{Q}_{\mathrm{A}}^{\mathrm{opt}} = \mathbf{V}_{\mathrm{Q}} \mathbf{U}_{\mathrm{Q}}^\mathrm{H},
	\label{eq-uni}
\end{align} where $\mathbf{V}_{\mathrm{Q}}$ and $\mathbf{U}_{\mathrm{Q}}$ are defined  based on the following SVD
\begin{align}
{\bf{F}}_{\rm{A}}^\mathrm{H} \, [{\bf{V}}_{\boldsymbol{\mathcal{H}}}]_{:,1:L} {\mathbf{\Lambda}_\mathrm{A}} = \mathbf{U}_{\mathrm{Q}} \mathbf{\Sigma}_{\mathrm{Q}} \mathbf{V}_{\mathrm{Q}}^\mathrm{H}.
\end{align}

Now that for given  $ {\bf Q}_{\rm A} $ and $\mathbf{\Lambda}_\mathrm{A} $, the optimal analog precoder $ \mathrm{\mathbf{F}_A} $ is  \cite{HanzoHybridTCOMM2016} \begin{align}
\label{phase_projection}
{\bf F}_{\rm A}^{\rm opt} = {\bf P}_{\mathcal F} \left( [ {\bf{V}}_{\boldsymbol{\mathcal{H}}}]_{:,1:L} {\mathrm{\mathbf{\Lambda}_A}} {\bf{Q}}_{\rm{A}} \right),
\end{align}
where the phase projection ${\bf P}_{\mathcal F}( \mathrm{\mathbf{A}} )$ is defined as
\begin{align}
	\left[ {\bf P}_{\mathcal F}( \mathrm{\mathbf{A}} ) \right]_{i,j} =
	\begin{cases}
	\left[ \mathrm{\mathbf{A}} \right]_{i,j} / | \left[ \mathrm{\mathbf{A}} \right]_{i,j} |, & \text{if} \  \left[ \mathrm{\mathbf{A}} \right]_{i,j} \neq 0 \\
	1, & \text{otherwise}.
	\end{cases}
\end{align}
Using (\ref{eq-diagonal-matrix}), (\ref{eq-uni}) and (\ref{phase_projection}), the phased projection based analog precoder optimization is proposed in Algorithm~\ref{alg-analog-design}.

\begin{algorithm}[t]
	\caption{Analog Precoder Design}
	\label{alg-analog-design}
	\begin{algorithmic}[1]
		\REQUIRE{ Left singular matrix of equivalent channel $ [{\bf{V}}_{\boldsymbol{\mathcal{H}}}]_{:,1:L} $, algorithm threshold $ \zeta $. }
		\STATE{ Initialize ${\bf{F}}_{\rm{A}}$ with $\mathbf{P}_{\mathcal{F}} \{ [{\bf{V}}_{\boldsymbol{\mathcal{H}}}]_{:,1:L} \} $. }
		\WHILE{ the decrement of the objective function in  (\ref{eq-pre-org}) is larger than $ \zeta $}
		\STATE{ Calculate  ${\mathbf{\Lambda}_{\rm A}}$ based on (\ref{eq-diagonal-matrix}). }
		\STATE{ Calculate  $\mathbf{Q}_{\mathrm{A}} $ based on (\ref{eq-uni}). }
		\STATE{ Calculate $\mathbf{F}_{\mathrm{A}}$ based on (\ref{phase_projection}). }
		\STATE{ Update the decrement value of the objective function in (\ref{eq-pre-org}). }
		\ENDWHILE
		\RETURN{$ \mathbf{F}_{\mathrm{A}} $.}
	\end{algorithmic}
\end{algorithm}

\noindent \textbf{Analog Processor Design}

Based on Conclusion 3, the optimal structure of analog processor is similar to the analog precoder, but a bit complicated in that the noise variance is tangled in the analog processor formulation. In this case, the left singular matrix of $ {\bf{R}}_{\rm{n}}^{1/2} {\bf{G}}_{\rm{A}}^{\rm{H}} $ is required to match the first $ L $ column of left singular matrix of effective channel, i.e., $ [{\bf{U}}_{\boldsymbol{\mathcal{H}}}]_{:,1:L} $. \begin{algorithm}[ht]
	\caption{Iterative Analog Processor Design}
	\label{alg-unit-mod}
	\begin{algorithmic}[1]
		\REQUIRE{ The matrix $ [{\bf{U}}_{\boldsymbol{\mathcal{H}}}]_{:,1:L} $, the unitary matrix $ \mathbf{Q}_{\mathrm{G}} $, the diagonal matrix $ \mathbf{\Lambda}_{\mathrm{G}} $, controlling factor $ \eta $, and convergent threshold $ \upsilon $. }
		\STATE{ Compute $ \mathbf{W} $ in (\ref{QCQP_3}). }
		\STATE{ Initialize constant-modulus processor as ${\mathbf{r}}_{(0)} = \frac{\sqrt{2}}{2}\mathbf{1} $. }
		\WHILE{The decrement of the objective function in (\ref{QCQP_3}) is larger than $ \upsilon $}
		\STATE{ Calculate $ \mathbf{P} $ using (\ref{para-tangent}) based on  $ \mathbf{G}_{\mathrm{A}} $ computed in the previous iteration. }
		\STATE{ Find out the optimal solution of (\ref{QCQP_3}) based on (\ref{eq-opt-QP}). }
		\STATE{ Update the decrement of the objective function in (\ref{QCQP_3}). }
		\ENDWHILE
		\STATE{ Construct  $ \mathbf{G}_{\mathrm{A}} $ based on the  optimal solution in Step 5. }
		\RETURN{ $ {\mathbf{G}}_{\mathrm{A}} $. }
	\end{algorithmic}
\end{algorithm}Thus, analogous to the analog precoder design in (\ref{eq-pre-org}), we have the following optimization problem
\begin{align}
	\min_{{\bf{G}}_{\rm{A}},{\mathrm{\mathbf{\Lambda}_G}},{\bf{Q}}_{\rm{G}}} \;\; &  \big\Vert [{\bf{U}}_{\boldsymbol{\mathcal{H}}}]_{:,1:L}
	{\mathbf{\Lambda}_\mathrm{G}} {\bf{Q}}_{\rm{G}} - {\bf{R}}_{\rm{n}}^{1/2} {\bf{G}}_{\rm{A}}^\mathrm{H} \big\Vert_{\rm{F}}^2
	\nonumber \\
	{\rm{s.t.}} \quad\;\; & {\bf{Q}}_{\rm{G}}{\bf{Q}}_{\rm{G}}^{\rm{H}} = {\bf{I}}\nonumber \\
	& {\bf{G}}_{\rm{A}} \in \mathcal{G}.
	\label{eq-comb-org}
\end{align}
The optimization of unitary matrix $ {\bf{Q}}_{\rm{G}} $ and diagonal matrix $ {\mathbf{\Lambda}_\mathrm{G}} $ in (\ref{eq-comb-org}) is exactly the same as that for the analog precoder optimization. However,  the optimization of the analog processor, ${\bf{G}}_{\rm{A}}$, in (\ref{eq-comb-org}) is different.

When noises from different antennas are correlated the analog processor design is more challenging than the analog precoder design. In order to overcome this challenge, problem (\ref{eq-comb-org}) is relaxed to minimize an upper bound of the original objective function. Applying
\begin{align}
	& \big\Vert [{\bf{U}}_{\boldsymbol{\mathcal{H}}}]_{:,1:L}
	{\mathbf{\Lambda}_\mathrm{G}} {\bf{Q}}_{\rm{G}} - {\bf{R}}_{\rm{n}}^{1/2} {\bf{G}}_{\rm{A}}^\mathrm{H} \big\Vert_{\mathrm{F}}^2 \notag \\
	\le \;\; & \lambda_{\mathrm{max}}( \mathbf{R}_{\mathrm{n}} )\big\Vert {\bf{R}}_{\rm{n}}^{-1/2} [{\bf{U}}_{\boldsymbol{\mathcal{H}}}]_{:,1:L}
	{\mathbf{\Lambda}_\mathrm{G}} {\bf{Q}}_{\rm{G}} - {\bf{G}}_{\rm{A}}^\mathrm{H} \big\Vert_{\mathrm{F}}^2,
\end{align}the objective function of (\ref{eq-comb-org}) is relaxed with $ \big\Vert {\bf{R}}_{\rm{n}}^{-1/2} [{\bf{U}}_{\boldsymbol{\mathcal{H}}}]_{:,1:L} {\mathbf{\Lambda}_\mathrm{G}} {\bf{Q}}_{\rm{G}} - {\bf{G}}_{\rm{A}}^\mathrm{H} \big\Vert_{\mathrm{F}}^2$. Note that solving (\ref{eq-comb-org}) is the same as that for the analog precoder design. It is obvious that this relaxation is tight when $ {\bf R}_{\rm n} = \sigma_n^2 {\bf I} $.

This relaxation may result in some performance loss. Inspired by the work in \cite{tractableTSP2017}, an iterative algorithm is also proposed to compute ${\bf{G}}_{\rm{A}}$. The constant modulus constraints is asymptotically satisfied via iteratively updating an additional constraint. This iterative algorithm is given in Algorithm~\ref{alg-unit-mod}, and detailed derivation is given in Appendix~\ref{Appendix_C}.

\subsection{Random Algorithm}

The proposed phase projection based analog transceiver design suffers from high computation complexity. This may prohibit the proposed analog transceiver design from practical implementation. In order to reduce complexity, we can
randomly generate analog precoder and processor matrices to avoid the heavy computations involved in the phase projection based algorithms.
In this random algorithm, we randomly select multiple matrices in the column or row space of ${\bf H}^{\rm{H}}$ and use their phase projections as the candidates for the analog transceiver design. Then the best candidate matrix is chosen according to some criterion.

Specifically, the random algorithm consists of three steps. First, a series of parameter matrices, denoted by $ \{\mathbf{R}_k\}$ and $\{{\mathbf{T}}_k\}$, are generated, whose elements are randomly generated following a specific distribution e.g., uniform distribution or Gaussian distribution. Secondly,  a series of candidate analog precoder and processor matrices are computed based on the parameter matrices. Specifically, based on the parameter matrices and after computing  ${\bf{H}}^{\rm{H}}{\bf{R}}_k$ and  ${\mathbf{T}}_k{\bf{H}}^{\rm{H}}$, the constant-modulus candidate matrices are obtained using their phase projections. Finally, the analog precoder and processor are chosen from these candidates according to the determinant of a certain matrix version SNR matrix. The procedure is detailed in Algorithm~\ref{alg-rand}.
\begin{algorithm}[t]
	\caption{Random Algorithm for Analog Transceiver Design}
	\label{alg-rand}
	\begin{algorithmic}[1]
		\REQUIRE{The number of transmitter antennas $ N $, number of RF-Chain $ L $, selection number $ K $, probability density function $ f_\mathrm{trans} (x) $, and ${\bf{H}}$ }
		\STATE{ Generate $ K $ parameter matrices, $ {\bf R}_{1},\ldots,{\bf R}_{K} \in \mathbb{C}^{ M \times L} $, whose elements are randomly generated based on $ f_\mathrm{trans} (x) $. }
		\STATE{ Rotate the channel as $ \mathrm{\mathbf{H}^H} {\bf R}_{k} $. Calculate ${\bf F}_{k} = \mathrm{\mathbf{P}}_{\mathcal{F}} \left( \mathrm{\mathbf{H}^H} {\bf R}_{k} \right) $. }
		\STATE{ ${\bf{F}}_{\rm{max}}={\arg \max}_{{\bf{F}}_i} \left\lvert { \bf F}_{i}^{\mathrm{H}} {\bf H}^{\mathrm{H}} \mathrm{ \mathbf{R}_n^{-1} } {\bf H}{\bf F}_{i} \right\rvert $. }
		\STATE{ Generate $ K $ parameter matrices $ {\bf T}_{1},\ldots,{\bf T}_{K} \in \mathbb{C}^{ L \times N } $ randomly based on $ f_\mathrm{trans} (x) $. }
		\STATE{ Rotate the channel as $ {\bf T}_{k} \mathrm{\mathbf{H}^H} $. Calculate $ {\bf G}_{k} = \mathrm{\mathbf{P}}_{\mathcal{F}} \left( {\bf T}_{k} \mathrm{\mathbf{H}^H} \right) $. }
		\STATE{  $ {\bf{G}}_{\max}={\arg \max}_{{\bf{G}}_i}\bigl\lvert {\bf G}_{i} \mathrm{ \mathbf{R}_n^{-1/2} } {\bf H} {\bf H}^{\mathrm{H}} \mathrm{ \mathbf{R}_n^{-1/2} } {\bf G}_{i}^{\mathrm{H}} \bigr\rvert $.}
		\RETURN{ $ {\bf F}_\mathrm{A} = {\bf F}_\mathrm{max} $, $ {\bf G}_\mathrm{A} = {\bf G}_\mathrm{max} $ }
	\end{algorithmic}
\end{algorithm}


\section{Simulation Results}

In this part, some numerical results are provided to assess the performance of the proposed hybrid transceiver design. As our algorithms are applicable to any frequency band, both microwave frequency band and mmWave frequency band are simulated.
In addition, quantization of phase shifters is also taken into account.
\begin{figure}[!t]
	\centering
	\includegraphics[width = .4\textwidth, trim={0 1 1 1}, clip]{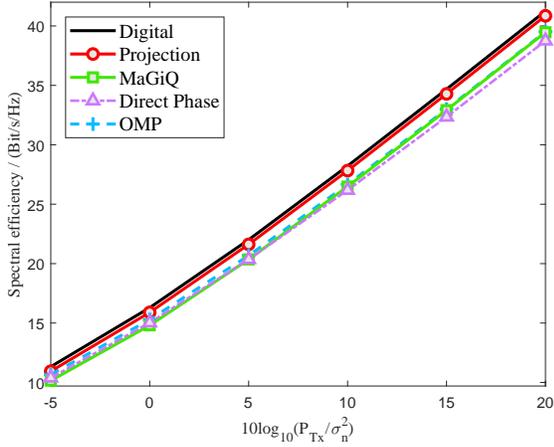}
	\caption{Spectral efficiency comparison of 5 different hybrid transceiver design methods. Here, $ 32 \times 16 $ mmWave channel model is adopted in the simulation. Both the transmitter and receiver are equipped with $ L = 4 $ RF-chains and the system is conveying $ D = 4 $ data streams.}
	\label{fig20}
\end{figure}
More specifically, both mmWave channel model $ {\mathrm{\mathbf{H}_{m}}} $ and classic Rayleigh channel model $ {\mathrm{\mathbf{H}_{r}}} $ are tested. For mmWave channel, $ {\mathrm{\mathbf{H}_{m}}} $, the uniformed linear arrays (ULA) is adopted.
 Unless otherwise specified, it is assumed that 1) the mmWave channel has $ N_{\rm cl} = 2 $ clusters with each of them containing $ N_\mathrm{path} = 5 $ paths; 2) the azimuth angle spread of transmitter is restricted to $ 7.5^{\circ} $ at the mean of azimuth angle $ \hat{\theta} = 45^{\circ} $, and the receiver is omni-directional; 3) the path loss factors obey the standard Gaussian distribution; 4) the inter-antenna spacing $ d $ equals to half-wavelength. The channel is normalized to meet $ \mathbb{E}{ \left\lbrace \lVert {\bf H}_{\rm m} \rVert_{\rm{F}}^2 \right\rbrace } = NM $.
For the random phase algorithm, we set $ K = 10 $, which means that the best analog precoder and processor are selected from 10 candidates and  uniform distribution is utilized, i.e., $ f_{\mathrm{ trans }}(x) = 1$ for $0 \le x \le 1 $. We average the result over 2,000 independent trials. The transmitting power is denoted as $ P_{\mathrm{Tx}} $.  OMP and MaGiQ algorithms refers to the corresponding algorithms in \cite{HeathTWC2014} and \cite{Y.C.Eldar2017}, respectively. The analog precoder and processor for the direct phase algorithm are obtained by phase projection.

\begin{figure}[!t]
	\centering
	\includegraphics[width = .4\textwidth, trim={0 1 0 0}, clip]{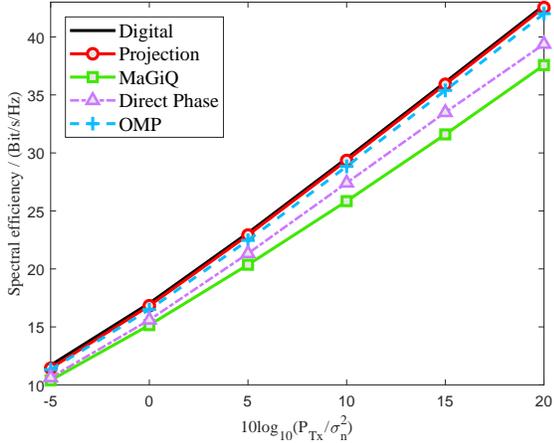}
	\caption{Spectral efficiency comparison of 5 different hybrid transceiver design methods. The $ 32 \times 16 $ mmWave channel model, which involves $ N_{\rm cl} = 3 $ clusters with $ N_{\rm path} = 5 $ multipath at each cluster, is adopted in the simulation. Both the transmitter and receiver are equipped with $ L = 6 $ RF-chains and the system is conveying $ D = 4 $ data streams.}
	\label{fig21}
\end{figure}
\begin{figure}[!ht]
	\centering
	\includegraphics[width = 0.4\textwidth, trim={0 0 0 0}, clip]{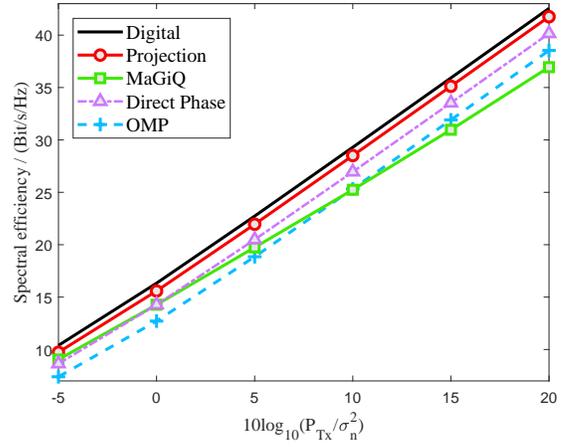}
	\caption{Spectral efficiency comparison of 5 different hybrid transceiver design methods. The $ 32 \times 16 $ Rayleigh channel model is adopted in the simulation. Both the transmitter and receiver are equipped with $ L = 6 $ RF-chains and the system is conveying $ D = 4 $ data streams.}
	\label{fig22}
\end{figure}

Fig.~\ref{fig20} demonstrates spectral efficiency versus the transmit power for different algorithms, where the hybrid transceiver is with $ N = 32 $ transmit antennas, $ M = 16 $ receive antennas, and 4 data streams. Both the transmitter and receiver are equipped with $ L = 4 $ RF-chains.
From Fig.~\ref{fig20}, the proposed phased  projection based hybrid transceiver design outperforms the other hybrid transceiver design algorithms. The performance of the proposed algorithm is very close to the full digital one.

Fig.~\ref{fig21} shows the performance of the hybrid transceiver design with 6 RF-chains for channel with $ N_{\rm cl} = 3 $ clusters, each with $ N_{\rm path} = 5 $ paths. From this figure, the proposed phase projection algorithm works well for different numbers of RF-chains and performs very close to the full-digital one and it is better than that of other hybrid transceiver designs. It is worth noting that the direct phase projection method performs\begin{figure}[!ht]
	\centering
	\includegraphics[width = .4\textwidth, trim={0 0 0 0}, clip]{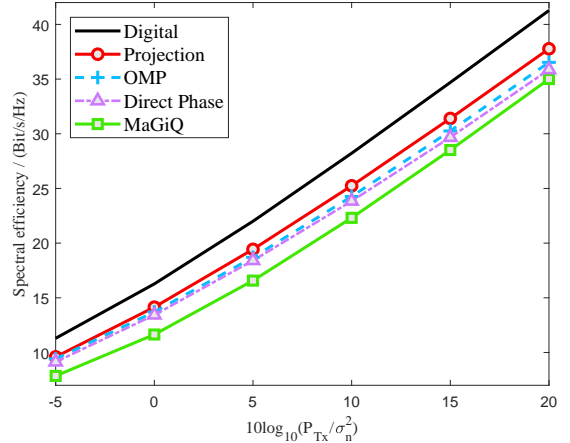}
	\caption{Spectral efficiency comparison of 5 different hybrid transceiver design methods concerning 2 bits quantization of phase shifters. The $ 32 \times 16 $ mmWave channel model is adopted in the simulation. Both the transmitter and receiver are equipped with $ L = 4 $ RF-chains and the system is conveying $ D = 4 $ data streams.}
	\label{fig23}
\end{figure} even better than OMP and MaGiQ. This is because the error bound of the method decreases when the number of RF chains increases \cite{HanzoHybridTCOMM2016}. However, as the limitation that the number of data streams should be equal to that of RF-chains \cite{Y.C.Eldar2017} is not satisfied in this case, MaGiQ algorithm is the worst at high SNR.

The following simulations focus on Rayleigh channels at micro-wave bands. Under this circumstance, the $ 32 \times 16 $ system is adopted with $ L = 6 $ RF-chains are in use transferring $ D = 4 $ data streams. After performing extensive simulation compared with randomly generated codebooks or DFT codebook, we found that the codebook constructed by the phase projection, i.e., $ \mathcal{C} = {\bf P}_{\mathcal{F}}( \mathrm{\mathbf{H}}) $, has much better performance. This codebook is used for performance comparison in the following simulation.

Fig.~\ref{fig22} compares the performance for the  different algorithms under Rayleigh channels. \begin{figure}[!ht]
	\centering
	\includegraphics[width = .4\textwidth, trim={20 0 30 20}, clip]{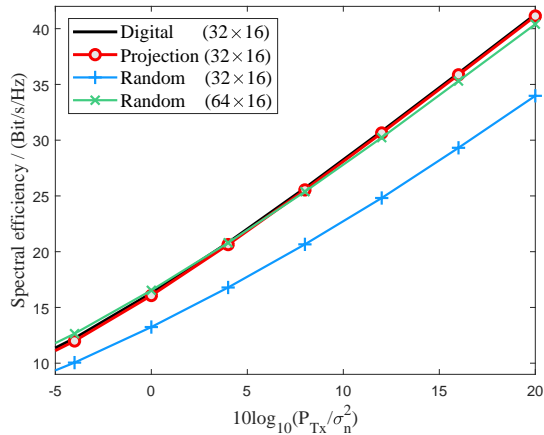}
	\caption{Spectral efficiency of random algorithm under mmWave channel. $ L = 6 $ RF-chains are assumed to be equipped both transmitter and receiver. Both $ 32 \times 16 $ and $ 64 \times 16 $ system are involved during the simulation. The number of data streams is set to be $ D = 4 $.}
	\label{fig24}
\end{figure} From the figure, the proposed algorithm obtains nearly the optimal performance as the full-digital one. The proposed algorithm performs better than MaGiQ algorithm in \cite{Y.C.Eldar2017}. Moreover, it is worth noting that even with the carefully chosen codebook, the OMP algorithm exhibits a large performance gap compared with the full-digital one, which indicates that the OMP algorithm is not suitable for micro-wave frequency bands.

As the practical analog phase shifters are often implemented by digital controller with finite resolution, Fig.~\ref{fig23} compares the performance of different hybrid transceiver designs for $ 32 \times 16 $ mmWave channel when phase quantization is taken into account. Each hybrid transceiver design only uses the phase shifter with 2-bit resolution and $ L = 4 $. From the figure the performance of the proposed hybrid transceiver design  still outperforms other hybrid transceiver designs with finite resolution phase shifters.

In Fig.~\ref{fig24}, both $ 32 \times 16 $ and $ 64 \times 16 $ mmWave channels are used to assess the performance. In this case, the number of RF-chains is $ 6 $. From Fig.~\ref{fig24}, with the same number of transmit antennas, the random algorithm is worse than that of the phase projection based algorithm. Although the random algorithm suffers nearly $ 5\,{\rm{dB}} $ performance loss comparing with the full-digital one, by involving more antennas at base station, e.g., $ N = 64 $, the performance of random algorithm will be comparable to the performance corresponding to the full-digital transmitter with $ 32 $ antennas. This implies that we can obtain appropriate performance using the low complexity random algorithm by simply increasing the number of transmit antennas. Because of its low complexity, the random algorithm will be a friendly algorithm for hardware realization.

\begin{figure}[ht]
	\centering
	\includegraphics[width = .4\textwidth, trim={20 0 20 10}, clip]{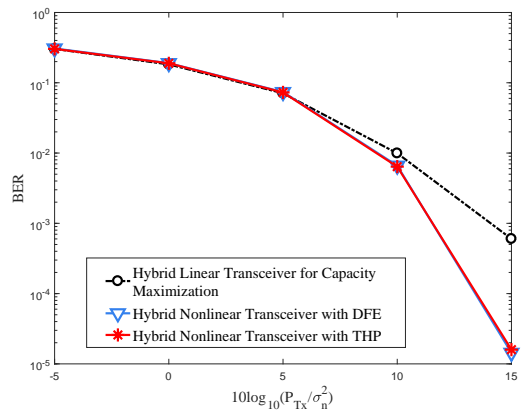}
	\caption{BERs of the linear hybrid transceiver for capacity maximization, nonlinear transceiver with DFD and nonlinear transceiver THP. The $ 32 \times 16 $ Rayleigh channel model is involved in the simulation. Both transmitter and receiver are equipped with $ L = 4 $ RF-chains transferring $ D = 4 $ data streams simultaneously. }
	\label{fig25}
\end{figure}

Fig.~\ref{fig25} shows the BER performances of different kinds of hybrid MIMO transceiver designs for $ 32 \times 16 $ Raleigh channel with 4 RF chains. In this case, there are 4 data streams and 16-QAM is used. From this figure, at high SNR, the BER performance of the hybrid nonlinear transceiver design is much better than that of the hybrid linear transceiver design. Furthermore, the hybrid nonlinear transceivers with THP and DFD have almost the same BER performance because of the duality between precoder design and processor design.

\section{Conclusions}
In this paper, we have investigated the hybrid digital and analog transceiver design for MIMO system based on matrix-monotonic optimization theory. We have proposed a unified framework for both linear and nonlinear transceivers.
Based on the matrix-monotonic optimization theory, the optimal transceiver structure for various MIMO transceivers has been derived, from which the function of analog transceiver part can be regarded as eigenchannel selection.
 Using the derived optimal structure, effective algorithms have been proposed considering the constant-modulus constraint.
Finally, it is shown that the proposed algorithms outperform existing hybrid transceiver designs.

\appendices
\section{Preliminary Definition of Majorization Theory}
\label{Appendix_A}
In this appendix, some fundamental functions in majorization theory are defined for the convenience of unified framework analysis. These definitions are also given in \cite{XingTSP201502} and in order to make the paper self-contained, they are also given here.

\textit{\textbf{Definition 1 \cite{inequalityMajorBook}: }}
For a $K\times 1$ vector $ {\bf x} \in \mathbb{R}^{K} $, the $ \ell $th largest element of $ {\bf x} $ is denoted as $ {x}_{[\ell]} $, and in other words, we have $ {x}_{[1]} \ge {x}_{[2]} \ge \cdots \ge {x}_{[K]} $. Based on this definition, for two vectors $ {\bf x}, {\bf y} \in \mathbb{R}^{K} $, it state that $ {\bf y} $ majorizes $ {\bf x} $ additively, denoted by $ {\bf x} \prec_{+} {\bf y} $, if and only if the following properties are satisfied
\begin{align}
\sum_{k = 1}^{p} {x}_{[k]} \le \sum_{k = 1}^{p} {y}_{[k]}, \; p = 1,2,\ldots, K-1 \; \text{and} \; \sum_{k = 1}^{K} {x}_{[k]} = \sum_{k = 1}^{K} {y}_{[k]}.
\end{align}

\textit{\textbf{Definition 2 \cite{inequalityMajorBook}: }}
A function $ f(\cdot) $ is Schur-convex if and only if it satisfies the following property
\begin{align}
{\bf x} \prec_{+} {\bf y} \, \Longrightarrow \, f( {\bf x} ) \le f( {\bf y} ).
\end{align}
On the other hand, a function $ f(\cdot) $ is  additively Schur-concave if $ -f(\cdot) $ is additively Schur-convex.

\textit{\textbf{Definition 3 \cite{XingTSP201502}: }}
For two $K\times 1$ vectors $ {\bf x}, {\bf y} \in \mathbb{R}^{K} $ with nonnegative elements, it states that the vector $ {\bf y} $ majorizes vector $ {\bf x} $ multiplicatively, i.e., $ {\bf x} \prec_{\times} {\bf y} $, if and only if the following properties are satisfied
\begin{align}
\prod_{k = 1}^{p} {x}_{[k]} \le \prod_{k = 1}^{p} {y}_{[k]}, \; p = 1,2,\ldots, K-1 \; \text{and} \; \prod_{k = 1}^{K} {x}_{[k]} = \prod_{k = 1}^{K} {y}_{[k]}.
\end{align}

\textit{\textbf{Definition 4 \cite{XingTSP201502}: }}
A function $ f(\cdot) $ is multiplicatively Schur-convex if and only if it satisfies the following property
\begin{align}
{\bf x} \prec_{+} {\bf y} \, \Longrightarrow \, f( {\bf x} ) \le f( {\bf y} ).
\end{align}
On the other hand, a function $ f(\cdot) $ is multiplicatively Schur-concave if $ -f(\cdot) $ is multiplicatively Schur-convex.

\section{The optimal ${\bf{B}}$}
\label{Appendix_Optimal_B}

Note that this optimal ${\bf{B}}$ for nonlinear transceiver was previously obtained in \cite{XingJSAC2012} when function  belongs to the family of multiplicatively Schur-concave/convex functions defined in Appendix A. The following presents a slightly different proof of the optimal B, which generalizes the result to the case with an arbitrary monotone increasing function $f(\cdot)$. Here, the function f operates only on the diagonal elements of ${\boldsymbol{\Phi}}_{\rm{MSE}}
	({\bf{G}}_{\rm{A}}, {\bf{F}}_{\rm{A}}, {\bf{F}}_{\rm{D}},{\bf{B}})$ and ${\bf{B}}$ is restricted as a strictly lower triangular matrix which specifies the use of nonlinear transceiver.

Based on the Cholesky decomposition
\begin{align}
{\boldsymbol{\Phi}}_{\rm{MSE}}^{\rm{L}}
	({\bf{G}}_{\rm{A}}, {\bf{F}}_{\rm{A}}, {\bf{F}}_{\rm{D}}) ={\bf{L}}{\bf{L}}^{\rm{H}},
\end{align}
we have
\begin{align}
	&{\boldsymbol{\Phi}}_{\rm{MSE}}
	({\bf{G}}_{\rm{A}}, {\bf{F}}_{\rm{A}}, {\bf{F}}_{\rm{D}}, {\bf{B}})\nonumber \\
	&= ({\bf{I}} + {\bf{B}}) {\boldsymbol{\Phi}}_{\rm{MSE}}^{\rm{L}}
	({\bf{G}}_{\rm{A}}, {\bf{F}}_{\rm{A}}, {\bf{F}}_{\rm{D}}) ({\bf{I}} + {\bf{B}})^{\rm H}\nonumber \\
&=({\bf{I}} + {\bf{B}}){\bf{L}}{\bf{L}}^{\rm{H}}({\bf{I}} + {\bf{B}})^{\rm H},
\end{align} based on which the $n$th diagonal element of ${\boldsymbol{\Phi}}_{\rm{MSE}}
	({\bf{G}}_{\rm{A}}, {\bf{F}}_{\rm{A}}, {\bf{F}}_{\rm{D}}, {\bf{B}})$ equals
\begin{align}
\label{app_1}
[{\boldsymbol{\Phi}}_{\rm{MSE}}
	({\bf{G}}_{\rm{A}}, {\bf{F}}_{\rm{A}}, {\bf{F}}_{\rm{D}}, {\bf{B}})]_{n,n}&=[({\bf{I}} + {\bf{B}}){\bf{L}}]_{n,:}[({\bf{I}} + {\bf{B}}){\bf{L}}]_{n,:}^{\rm{H}}\nonumber \\
&=\|[({\bf{I}} + {\bf{B}}){\bf{L}}]_{n,:}\|^2.
\end{align}In addition, as ${\bf{B}}$ is strictly lower triangular  it can be calculated that the last element of the vector $[({\bf{I}} + {\bf{B}}){\bf{L}}]_{n,:}$ equals $[{\bf{L}}]_{n,n}$, i.e.,
\begin{align}
\label{app_2}
[({\bf{I}} + {\bf{B}}){\bf{L}}]_{n,:}=[\cdots,[{\bf{L}}]_{n,n}].
\end{align} Therefore, from (\ref{app_1}) to (\ref{app_2})  the following relationship holds
\begin{align}
[{\boldsymbol{\Phi}}_{\rm{MSE}}
	({\bf{G}}_{\rm{A}}, {\bf{F}}_{\rm{A}}, {\bf{F}}_{\rm{D}}, {\bf{B}})]_{n,n}
&=\|[({\bf{I}} + {\bf{B}}){\bf{L}}]_{n,:}\|^2\ge [{\bf{L}}]_{n,n}^2.
\end{align}It is obvious that the above inequality can be achieved with equality as $[{\boldsymbol{\Phi}}_{\rm{MSE}}
	({\bf{G}}_{\rm{A}}, {\bf{F}}_{\rm{A}}, {\bf{F}}_{\rm{D}}, {\bf{B}})]_{n,n}
= [{\bf{L}}]_{n,n}^2$ when the following equality holds for different $n$
\begin{align}
({\bf{I}}+{\bf{B}}){\bf{L}} ={\rm{diag}}\{[[{\bf{L}}]_{1,1},\cdots,
[{\bf{L}}]_{L,L}]^{\rm{T}}\},
\end{align}based on which the optimal ${\bf{B}}$ equals
\begin{align}
{\bf{B}}^{\rm{opt}} ={\rm{diag}}\{[[{\bf{L}}]_{1,1},\cdots,
[{\bf{L}}]_{L,L}]^{\rm{T}}\}
{\bf{L}}^{-1}-{\bf{I}}.
\end{align}

\section{Fundamental Matrix Inequalities}
\label{Appendix_Inequality}
In this appendix, two fundamental matrix inequalities are given.
For two positive semi-definite matrices ${\boldsymbol X}$ and ${\boldsymbol Y}$, there are following EVDs defined
\begin{align}
	{\boldsymbol X}&={\bf{U}}_{\boldsymbol X} {\boldsymbol\Lambda}_{\boldsymbol X} {\bf{U}}^{\rm{H}}_{\boldsymbol X} \ \ \text{with} \ \ {\boldsymbol\Lambda}_{\boldsymbol X}\searrow \nonumber \\
	{\boldsymbol Y}&={\bf{U}}_{\boldsymbol Y} {\boldsymbol\Lambda}_{\boldsymbol Y} {\bf{U}}^{\rm{H}}_{\boldsymbol Y} \ \ \text{with} \ \ {\boldsymbol\Lambda}_{\boldsymbol Y}\searrow \nonumber \\
	{\boldsymbol Y}&={\bf{\bar U}}_{\boldsymbol Y}{\boldsymbol{\bar \Lambda}}_{\boldsymbol Y} {\bf{\bar U}}^{\rm{H}}_{\boldsymbol Y} \ \ \text{with} \ \ {\boldsymbol{\bar \Lambda}}_{\boldsymbol Y} \nearrow.
\end{align}
For the trace of the two matrices, we have the following fundamental matrix inequalities \cite{XingTSP201501}
\begin{align}
	&\sum_{i=1}^{N}\lambda_{i-1+N}({\boldsymbol X}) \lambda_i({\boldsymbol Y})\le {\rm{Tr}}({\boldsymbol X}{\boldsymbol Y})  \le \sum_{i=1}^{N}\lambda_i({\boldsymbol X}) \lambda_i({\boldsymbol Y}),
\end{align}
where $ \lambda_i( { \mathbf{X} } ) $ is the $ i $th ordered eigenvalue of $ { \mathbf{X} } $, and the left equality holds when ${\bf{U}}_{\boldsymbol X}={\bf{\bar U}}_{\boldsymbol Y}$. On the other hand, the right equality holds when ${\bf{U}}_{\boldsymbol X}={\bf{ U}}_{\boldsymbol Y}$.

\section{Optimal Structure of Analog Transceiver}
\label{Appendix_Analog_Transceiver}

It is worth noting that the nonzero singular values of the matrix, $ {\boldsymbol{\Pi}}_{\rm{R}} = {\bf{F}}_{\rm{A}} ({\bf{F}}_{\rm{A}}^{\rm{H}}{\bf{F}}_{\rm{A}})^{-\frac{1}{2}} $, are all ones. Similarly for ${\boldsymbol{\Pi}}_{\rm{L}}=
({\bf{G}}_{\rm{A}}{\bf{R}}_{\rm{n}}{\bf{G}}_{\rm{A}}^{\rm{H}})^{-1/2}
{\bf{G}}_{\rm{A}}{\bf{R}}_{\rm{n}}^{1/2}$,  the nonzero singular values of ${\boldsymbol{\Pi}}_{\rm{L}}$ are all ones.
  It implies that the singular values of ${\bf{F}}_{\rm{A}}$ and ${\bf{G}}_{\rm{A}}{\bf{R}}_{\rm{n}}^{1/2}$  do not affect the optimization problem. Based on the SVDs ${\bf{R}}_{\rm{n}}^{-1/2}{\bf{H}}={\bf{U}}_{{\boldsymbol{\mathcal{H}}}}
	{\boldsymbol \Lambda}_{\boldsymbol{\mathcal{H}}}
{\bf{V}}_{\boldsymbol{\mathcal{H}}}^{\rm{H}}
$, ${\bf{R}}_{\rm{n}}^{1/2} {\bf{G}}_{\rm{A}}^{\rm{H}} = {\bf{U}}_{{\bf{R}}{\bf{G}}} {\boldsymbol{\Lambda}}_{{\bf{R}}{\bf{G}}}
{\bf{V}}_{{\bf{R}}{\bf{G}}}^{\rm{H}}
 $, and ${\bf{F}}_{\rm{A}}={\bf{U}}_{{\bf{F}}_{\rm{A}}} {\boldsymbol{\Lambda}}_{{\bf{F}}_{\rm{A}}}{\bf{V}}_{{\bf{F}}_{\rm{A}}}^{\rm{H}}$
with the singular values in decreasing order,
the objective function in (\ref{eq-analog-problem}) becomes
\begin{align}
\label{New_Matrix_Objective}
{\bf{\tilde F}}_{\rm{D}}^{\rm{H}}{\bf{V}}_{{\bf{F}}_{\rm{A}}}
{\boldsymbol \Lambda}_{\rm{R}}^{\rm{T}}
{\bf{U}}_{{\bf{F}}_{\rm{A}}}^{\rm{H}}{\bf{H}}^{\rm{H}}
{\bf{U}}_{\rm{RG}}{\boldsymbol \Lambda}_{\rm{L}}^{\rm{T}}
{\boldsymbol \Lambda}_{\rm{L}}{\bf{U}}_{\rm{RG}}^{\rm{H}}
{\bf{H}}{\bf{U}}_{{\bf{F}}_{\rm{A}}}{\boldsymbol \Lambda}_{\rm{R}}{\bf{V}}_{{\bf{F}}_{\rm{A}}}^{\rm{H}}{\bf{\tilde F}}_{\rm{D}}
\end{align}where the diagonal elements of the diagonal matrices ${\boldsymbol \Lambda}_{\rm{R}}$ and  ${\boldsymbol \Lambda}_{\rm{L}}$ satisfies
\begin{align}
&[{\boldsymbol \Lambda}_{\rm{R}}]_{i,i}=1, \ \ i\le L \nonumber \\
& [{\boldsymbol \Lambda}_{\rm{R}}]_{i,i}=0, \ \ i> L  \nonumber \\
&[{\boldsymbol \Lambda}_{\rm{L}}]_{i,i}=1, \ \ i\le L \nonumber \\
& [{\boldsymbol \Lambda}_{\rm{R}}]_{i,i}=0, \ \ i> L.
\end{align}Therefore, ${\bf{F}}_{\rm{A}}$ and ${\bf{G}}_{\rm{A}}$ do not affect the optimal solution. Moreover, the unitary matrices ${\bf{V}}_{{\bf{F}}_{\rm{A}}}$ and ${\bf{V}}_{\rm{RG}}$ do not affect the optimal solution as ${\bf{\tilde F}}_{\rm{D}}$ in the constraint is unitary invariant.

Based on the above the discussion and (\ref{New_Matrix_Objective}), the remaining task to maximize the singular values of matrix $[{\bf{U}}_{\rm{RG}}^{\rm{H}}
{\bf{H}}{\bf{U}}_{{\bf{F}}_{\rm{A}}}]_{1:L,1:L}$. Note that ${\bf{U}}_{\rm{RG}}$ and ${\bf{U}}_{{\bf{F}}_{\rm{A}}}$ are unitary matrices, for the optimal solution, the left eigenvectors of its first $ L $ largest singular values of $ {\bf F}_{\rm A} $ should have the maximum inner product with $[{\bf{V}}_{\boldsymbol{\mathcal{H}}}]_{:,1:L}$ i.e.,
\begin{align} [{\bf{U}}_{{\bf{F}}_{\rm{A}}}]_{:,1:L}^{\rm{opt}}
={\rm{arg\,max}}
	\{
	\|[{\bf{V}}_{\boldsymbol{\mathcal{H}}}]_{:,1:L} [{\bf{U}}_{{\bf{F}}_{\rm{A}}}]_{:,1:L}^{\rm{H}}\|_{\rm{F}}^2
	\}.
\end{align}

Similarly for the optimal solution, the left eigenvectors of its first $ L $ largest singular values of $ {\bf{R}}_{\rm{n}}^{1/2} {\bf{G}}_{\rm{A}}^{\mathrm{H}} $ should have the maximum inner product with $[{\bf{U}}_{\boldsymbol{\mathcal{H}}}]_{:,1:L}$, i.e.,
\begin{align} [{\bf{U}}_{{\bf{R}}{\bf{G}}}]_{:,1:L}^{\rm{opt}}={\rm{arg\,max}}
\{\|[{\bf{U}}_{\boldsymbol{\mathcal{H}}}]_{:,1:L} [{\bf{U}}_{{\bf{R}}{\bf{G}}}]_{:,1:L}^{\rm{H}}\|_{\rm{F}}^2\}.
\end{align}

\section{Analog Transceiver Design}
\label{Appendix_C}

For fixed ${\mathbf{\Lambda}_\mathrm{G}}$ and ${\bf{Q}}_{\rm{G}}$, the optimization problem (\ref{eq-comb-org}) can be transferred into the following vector variable optimization problem
\begin{align}
\label{QCQP_1}
	\min_{\mathbf{r}} \;\; & {\mathbf{r}}^\T \mathbf{W} {\mathbf{r}} - \mathbf{p}^\T {\mathbf{r}} - {\mathbf{r}}^\T \mathbf{p} + q \notag \\
	\text{s.t.} \;\;\; & \mathbf{r}^\T \mathbf{K}_{i} \mathbf{r} = a^2, \quad i = 1,2,\ldots, NL.
\end{align}
The vector $ \mathbf{r} $ is constructed via vectorizing $ \mathbf{G}_{\mathrm{A}} $, i.e.,
\begin{align*}
	\mathbf{r} = \big[ \Re \{ \vecm{( \mathbf{G}_{\mathrm{A}} )} \}^\T, \, \Im \{ \vecm{( \mathbf{G}_{\mathrm{A}} )} \}^\T \big]^\T,
\end{align*}
and the matrices $ \mathbf{W}, \, \mathbf{K}_{i} $ and vector $ \mathbf{p} $ are defined as follows:
\begin{align}
	\notag \mathbf{W} & =
	\begin{bmatrix}
	\; \Re \{ \mathbf{I} \otimes { \mathbf{R}_{\mathrm{n}} } \} & - \Im \{ \mathbf{I} \otimes { \mathbf{R}_{\mathrm{n}} } \} \;\; \\
	\; \Im \{ \mathbf{I} \otimes { \mathbf{R}_{\mathrm{n}} } \} & \Re \{ \mathbf{I} \otimes { \mathbf{R}_{\mathrm{n}} } \} \;\;
	\end{bmatrix}, \\
	\notag \mathbf{K}_{i} & = \mathrm{diag} \Big\lbrace \bigl[ \mathbf{0}_{(i-1) \times 1}^\T, 1 ,\mathbf{0}_{(NL - 1) \times 1}^\T, 1 ,\mathbf{0}_{(NL - i) \times 1}^\T \bigr] \Big\rbrace,
\end{align}
and
\begin{align}
	\mathbf{p} =
	\begin{bmatrix}
	\Re \{ \big( \mathbf{I} \otimes { \mathbf{R}_{\mathrm{n}}^{ {1}/{2}} } \big)^{\mathrm{H}} \vecm{ \big( [{\bf{U}}_{\boldsymbol{\mathcal{H}}}]_{:,1:L} {\mathbf{\Lambda}_\mathrm{G}} {\bf{Q}}_{\rm{G}} \big) } \} \\
	\Im \{ \big( \mathbf{I} \otimes { \mathbf{R}_{\mathrm{n}}^{ {1}/{2}} } \big)^{\mathrm{H}} \vecm{ \big( [{\bf{U}}_{\boldsymbol{\mathcal{H}}}]_{:,1:L} {\mathbf{\Lambda}_\mathrm{G}} {\bf{Q}}_{\rm{G}} \big) } \}
	\end{bmatrix}.
\end{align}The constant scalar, $q$, in (\ref{QCQP_1}) equals $q = || \vecm{ \big( [{\bf{U}}_{\boldsymbol{\mathcal{H}}}]_{:,1:L} {\mathbf{\Lambda}_\mathrm{G}} {\bf{Q}}_{\rm{G}} \big) } ||_2^2$.

Note that because of the constant modulus constraints, the term ${\bf{r}}^{\rm{T}}{\bf{r}}$ is a constant. As a result, for a constant real scalar, $\eta$, the objective function in  (\ref{QCQP_1}) is equivalent to ${\mathbf{r}}^\T (\mathbf{W}+\eta{\bf{I}}) {\mathbf{r}} - \mathbf{p}^\T {\mathbf{r}} - {\mathbf{r}}^\T \mathbf{p} + q$. As the constant  modulus constraints in (\ref{QCQP_1}) are all quadratic equalities, the optimization problem (\ref{QCQP_1}) is nonconvex. Following the idea of \cite{tractableTSP2017}, an iterative algorithm is proposed via iteratively updating constraints to guarantee the constant modulus constraints. Specifically, at the $n$th iteration each constraint $\mathbf{r}^\T \mathbf{K}_{i} \mathbf{r} = a^2$ is replaced by $\mathbf{\tilde r}_{(n-1)}^\T \mathbf{K}_{i} \mathbf{r}_{(n)} = a^2$ where $\mathbf{\tilde r}_{(n-1)}$ is a vector computed based on ${\bf{r}}$ computed in the $(n-1)$th iteration. After stacking  $\mathbf{\tilde r}_{(n-1)}^\T \mathbf{K}_{i}$ for $i = 1,2,\ldots NL$ in ${\bf{P}}_{(n-1)}$, optimization problem (\ref{QCQP_1}) is transferred to
\begin{align}
\label{QCQP_3}
	\min_{\mathbf{r}_{(n)}} \;\; & {\mathbf{r}}_{(n)}^\T (\mathbf{W}+\eta{\bf{I}}) {\mathbf{r}}_{(n)} - \mathbf{p}^\T {\mathbf{r}}_{(n)} - {\mathbf{r}}_{(n)}^\T \mathbf{p} + q \notag \\
	\text{s.t.} \;\;\; & {\bf{P}}_{(n-1)}\mathbf{r}_{(n)} = a^2{\bf{1}},
\end{align}
where the matrix $ \mathbf{P}_{(n-1)} $ is defined as
\begin{align}
	&[ \mathbf{P}_{(n-1)} ]_{\ell,j} \nonumber \\
=&
	\begin{cases}
	\cos \big( \angle [{\rm{vec}}({\bf{G}}_{{\rm{A}},(n-1)})]_{\ell} \big) & \text{if } \ell = j, \; \ell \le NL \\
	\sin \big( \angle [ {\rm{vec}}({\bf{G}}_{{\rm{A}},(n-1)}) ]_{\ell} \big) & \text{if } j = \ell+NL, \; \ell \le NL \\
	0 & \text{otherwise.}
	\end{cases}
	\label{para-tangent}
\end{align}The vector $ \mathbf{1} $ is a column vector with all elements equal to 1. As proved in \cite{tractableTSP2017}, when $\eta \ge \sigma_{\max} NL / 8 + ||\mathbf{p}||_2^2$,
where $ \sigma_{\max} $ is the largest eigenvalue of $ { \mathbf{R}_{\mathrm{n}} }$,
the optimal solution of the iterative optimization (\ref{QCQP_3}) minimizes the objective function and satisfies the constant modulus constraints asymptotically. As (\ref{QCQP_3}) is convex at each iteration, based on its KKT conditions, at the $n$th iteration the optimal solution of (\ref{QCQP_3}) is
\begin{align}
{\mathbf{r}}_{(n)} =
	( {\mathbf{W}} + \eta \mathbf{I} )^{-1}\left({\bf{q}}+\frac{\lambda}{2}{\bf{P}}_{(n-1)}^{\rm{T}}\right)
	\label{eq-opt-QP}
\end{align}with
\begin{align}
\frac{\lambda}{2}&=\left(	{\bf{P}}_{(n-1)}( {\mathbf{W}} + \eta \mathbf{I} )^{-1}{\bf{P}}_{(n-1)}^{\rm{T}}\right)^{-1}\nonumber \\
& \ \ \ \ \ \times\left(a^2{\bf{1}}-{\bf{P}}_{(n-1)}( {\mathbf{W}} + \eta \mathbf{I} )^{-1}{\bf{q}}\right).
\end{align}
In a nutshell, the iterative algorithm is given in Algorithm~\ref{alg-unit-mod}. Using the iterative algorithm, the numerical result of analog processor can be found.

\bibliographystyle{IEEEtran}
\bibliography{IEEEabrv,Hybrid_Matrix_Monotonic_Optimization}

\end{document}